\documentclass[aps,prc,preprint,groupedaddress,showpacs]{revtex4}
\usepackage[dvipdfmx]{graphicx}
\usepackage{amsmath}
\usepackage{amssymb}

\begin{document}
\title{Single-phonon and multi-phonon excitations of the $\gamma$ vibration in rotating odd-$A$ nuclei}

\author{Masayuki Matsuzaki}
\email[]{matsuza@fukuoka-edu.ac.jp}
\affiliation{Department of Physics, Fukuoka University of Education, 
             Munakata, Fukuoka 811-4192, Japan}

\date{\today}

\begin{abstract}
\begin{description}
\item[Background] Collective motions in quantum many-body systems are described as 
bosonic excitations. Multi-phonon excitations in atomic nuclei, however, were observed 
very rarely. In particular, the first two-phonon $\gamma$ vibrational ($2\gamma$) 
excitation in odd-$A$ nuclei was reported in 2006 and only a few have been known 
so far. Two theoretical calculations for the data on $^{103}$Nb were performed, one of 
which was done by the present author within a limited model space up to $2\gamma$ basis states. 
Quite recently, conspicuously enhanced $B(E2)$s, reduced $E2$ transition probabilities, 
feeding $2\gamma$ states were observed 
in $^{105}$Nb and conjectured that their parent states, called band (4), are candidates 
of $3\gamma$ states.
\item[Purpose] In the present work, the model space is enlarged up to $4\gamma$ basis states. 
The purpose is twofold: One is to see how the description of $2\gamma$ eigenstates in the previous 
work is improved, and the other is to examine the existence of collective $3\gamma$ eigenstates, 
and when they exist, study their collectivity through calculating interband $B(E2)$s. 
\item[Method] The particle-vibration coupling model based on the cranking model and 
the random-phase approximation is used to calculate the vibrational states in rotating 
odd-$A$ nuclei. Interband $B(E2)$s are calculated by adopting the method of the 
generalized intensity relation. 
\item[Results] The present model reproduces well the energy spectra and $B(E2)$s of 
$0\gamma$ -- $2\gamma$ states in $^{103}$Nb and $^{105}$Nb. For $3\gamma$ states, calculated 
spectra indicate that the most collective state with the highest $K$ at zero rotation 
feels strong Coriolis force after rotation sets in and consequently is observed with 
lowered $K$, where $K$ is the projection of the angular momentum to the $z$ axis. 
The calculated states account for the observed enhanced $B(E2)$s within factors of 
2 -- 3. 
\item[Conclusions] The present calculation with the enlarged model space reproduces 
the observed $0\gamma$ -- $2\gamma$ states well and predicts properties of collective $3\gamma$ states. 
The most collective one is thought to be the main component of the observed band (4) 
from the analyses of the energy spectra and interband $B(E2)$s although some mixing 
with states that are not included in the present model would be possible. 
\end{description}
\end{abstract}

\maketitle

\section{Introduction}

Collective motions in quantum many-body systems are formed as coherent superpositions 
of many individual degrees of freedom, and are described as bosonic excitations. 
In atomic nuclei, one of finite many-body systems, the representative is vibrations of 
the surface of the average potential produced self-consistently to the nucleon distribution. 
However, repeated excitations, the characteristic 
of bosons, are not always observed. Even when observed, their strengths spread over many 
eigenstates because collective and individual noncollective excitations have similar 
energy scale. Therefore, existence and properties of multiple 
excitations have been a longstanding subject of theoretical and experimental studies. 

Famous examples are known in high-lying giant resonances. Double excitations of the 
one with the same multipole and of different types have been observed; see for example, 
a review~\cite{ABE}. Among low-lying vibrations, the multi-phonon quadrupole vibrational 
states in spherical nuclei have long been studied around $^{110}$Cd~\cite{BM,Cd110}. 
In axially symmetric or weakly triaxial deformed nuclei, the two-phonon $\gamma$ 
vibration, denoted as $2\gamma$ hereafter, was studied for three decades as 
concisely reviewed in Ref.~\cite{MM1}, but 
observed only in $^{168}$Er, $^{166}$Er, $^{164}$Dy, $^{232}$Th, $^{106}$Mo and 
$^{104}$Mo in even-even nuclei. Recently, another candidate was reported in the weakly 
deformed, $\gamma$-soft nucleus, $^{138}$Nd~\cite{Nd138}. 

In odd-$A$ nuclei, vibrations of the average potential alter the particle motion and 
the change thus caused affects back the average potential. Consequently the particle 
motion and the vibration couple to each other to various degrees. This coupling makes 
the excitation spectrum complex in general. From another point of view, however, 
a stretched parallel coupling of the high-$j$ particle and $K=2$ phonons can 
produce high-$K$ states that can hardly mix with other states with lower $K$. 
Here $j$ is the single-particle angular momentum, and $K=2$ the projection to the $z$ 
axis of the angular momentum carried by the $\gamma$ vibration. 
Accordingly there can be more opportunities to observe multi-phonon $\gamma$-vibrational 
states. 

Prior to experimental observations, numerical predictions for odd-$A$ nuclei were made 
in Ref.~\cite{DP}. The first observation was made 10 years later in $^{105}$Mo~\cite{Mo105}. 
Soon after this, similar bands were observed in $^{103}$Nb~\cite{Nb103} and in 
$^{107}$Tc~\cite{Tc107}. The first realistic theoretical calculation in terms of the 
triaxial projected shell model was reported for $^{103}$Nb~\cite{SBSP} in 2010. 
The calculation in terms of the particle-vibration coupling (PVC) model based on the cranking 
model and the random-phase approximation (RPA) was done in 2011~\cite{MM1}. 

As another type of multi-phonon state in strongly triaxial deformed nuclei, two-phonon 
wobbling bands were observed in $^{163}$Lu~\cite{Lu163} and $^{165}$Lu~\cite{Lu165} and analyzed 
in Ref.~\cite{MO}. Furthermore, the rotational bands built on the high-$K$ multi-quasiparticle 
states can be interpreted theoretically as a multi-phonon excitation of the precession 
mode, which is the axially-symmetric limit of the wobbling mode~\cite{SMM}. 

Quite recently, a $2\gamma$ band in $^{105}$Nb bearing much resemblance to that in $^{103}$Nb 
was reported~\cite{Nb105}. The characteristic feature common to these two isotopes but 
beyond the scope of Ref.~\cite{MM1} is that the observed band (4) is interpreted as a candidate 
of the $3\gamma$ band. In particular in $^{105}$Nb, $B(E2)$s from this band to the 
$2\gamma$ band are conspicuously enhanced compared with the Weisskopf unit. 

In Ref.~\cite{MM1}, the model, which was developed to study the signature dependence in $0\gamma$ 
bands in the rare-earth region~\cite{MSM,MM} and utilized later to study the 
two sequences with $K=\Omega\pm2$ of the $1\gamma$ band in $^{165}$Ho and the one with 
$K=\Omega+2$ in $^{167}$Er~\cite{Ge}, was applied to the $2\gamma$ band for the first time. 
Here $\Omega$ is the projection of the single-particle angular 
momentum to the $z$ axis. 
From a microscopic many-body theoretical point of view, the mechanism that leads to 
anharmonicity of the spectrum beyond the RPA was discussed in the studies of 
$2\gamma$ bands in even-even nuclei~\cite{MatsuoMatsu}, however, the present model 
concentrates on that peculiar to odd-$A$ systems. 
The result was that the $0\gamma$ and $1\gamma$ bands 
were reproduced perfectly but the calculated $2\gamma$ band was somewhat higher in energy 
than the observed one as in the other calculation~~\cite{SBSP}. 

In the present paper, the model space is enlarged up to $4\gamma$ basis states and the interband 
$B(E2)$s are also calculated utilizing the method of the generalized intensity relation 
(GIR)~\cite{BM,SN}. 
The purpose is twofold: One is to see how the description of $2\gamma$ states in Ref.~\cite{MM1} is 
improved, and the other is to look into the existence of $3\gamma$ states, and when they 
exist, study their collectivity through calculating vibrational $B(E2)$s. This is the first 
attempt to study realistic $3\gamma$ states in deformed nuclei, to the author's knowledge. 

Throughout this paper the $\hbar=1$ unit is used. 

\section{The model}

\subsection{Particle-vibration coupling}

Eigenstates of the odd-$A$ nucleus in a rotating frame are calculated in the particle-vibration 
coupling model based on the cranking model and the RPA. The adopted Hamiltonian is the same 
as in Ref.~\cite{MM1}, therefore recapitulated only briefly here. The detailed definitions of the 
adopted notations were given there~\cite{MM1}. 
I begin with the cranked Nilsson plus BCS one-body Hamiltonian, 
\begin{gather}
h'=h-\omega_\mathrm{rot}J_x , \\
h=h_\mathrm{Nil}-\Delta_\tau (P_\tau^\dagger+P_\tau)
                   -\lambda_\tau N_\tau , \\
h_\mathrm{Nil}=\frac{\mathbf{p}^2}{2M}
                +\frac{1}{2}M(\omega_x^2 x^2 + \omega_y^2 y^2 + \omega_z^2 z^2)
                +v_{ls} \mathbf{l\cdot s} 
                +v_{ll} (\mathbf{l}^2 - \langle\mathbf{l}^2\rangle_{N_\mathrm{osc}}) .
\end{gather}
This Hamiltonian gives 
quasiparticle states created by $a_\mu^\dagger$ and $a_{\bar\mu}^\dagger$ with signature 
$r=\exp{(-i\pi\alpha)}=-i$ and $+i$, respectively. Then I apply the RPA to the residual 
pairing plus doubly-stretched quadrupole-quadrupole interaction between quasiparticles. 
The interaction Hamiltonian is given by 
\begin{equation}
H_\mathrm{int}=
-\sum_{\tau=1,2} G_\tau \tilde P_\tau^\dagger \tilde P_\tau
-\frac{1}{2}\sum_{K=0,1,2} \kappa_K^{(+)} Q_K''^{(+)\dagger} Q_K''^{(+)} 
-\frac{1}{2}\sum_{K=1,2} \kappa_K^{(-)} Q_K''^{(-)\dagger} Q_K''^{(-)} ,
\end{equation}
where $Q_K''^{(\pm)}$ are the signature-coupled form of the quadrupole operators defined 
by the doubly-stretched coordinates. 

Among the RPA modes, $X_n^\dagger$, determined by $H_\mathrm{int}$, I choose the 
$\gamma$-vibrational phonons, $n=\gamma(\pm)$ with signature $r=\pm 1$, which have 
outstandingly large $K=2$ transition amplitudes. 
In terms of the quasiparticles and the $\gamma$-vibrational phonons thus 
determined, the particle-vibration coupling Hamiltonian takes the form
\begin{equation}
\begin{split}
H_\mathrm{couple}(\gamma)
     &={\sum_{\mu\nu}}\Lambda_{\gamma(+)}(\mu\nu)
       \left(X_{\gamma(+)}^\dagger a_\mu^\dagger a_\nu
           + X_{\gamma(+)}         a_\nu^\dagger a_\mu\right) \\
     &+\sum_{\mu\bar\nu}\Lambda_{\gamma(-)}(\mu\bar\nu)
       \left(X_{\gamma(-)}^\dagger a_\mu^\dagger a_{\bar\nu}
           + X_{\gamma(-)}         a_{\bar\nu}^\dagger a_\mu\right) \\
     &+\mbox{sig. conj.} 
\end{split}
\end{equation}
The coupling vertices are given by 
\begin{equation}
\begin{split}
   &\Lambda_{\gamma(+)}(\mu\nu)
             =-\sum_{K=0,1,2}\kappa_K^{(+)} T_K''^{(+)}
                             Q_K''^{(+)}(\mu\nu) ,\\
   &\Lambda_{\gamma(-)}(\mu\bar\nu)
             =-\sum_{K=1,2}\kappa_K^{(-)} T_K''^{(-)}
                             Q_K''^{(-)}(\mu\bar\nu) ,\\
   &\mbox{and sig. conj.} ,
\end{split}
\end{equation}
where $T_K''^{(\pm)}$ are the doubly-stretched quadrupole transition amplitudes 
associated with the $\gamma$-vibrational phonons and $Q_K''^{(\pm)}(\alpha\beta)$ 
denote quasiparticle scattering matrix elements. 

Eigenstates of the total Hamiltonian at each $\omega_\mathrm{rot}$ take the form 
\begin{equation}
\begin{split}
\left.|\chi_i\right\rangle 
&=\sum_{\mu}\psi_i^{(1)}(\mu)\left.a_\mu^\dagger|\phi\right\rangle \\
& +\sum_{\mu}\psi_i^{(3)}(\mu\gamma)\left.a_\mu^\dagger X_\gamma^\dagger|\phi\right\rangle
  +\sum_{\bar\mu}\psi_i^{(3)}(\bar\mu\bar\gamma)
\left.a_{\bar\mu}^\dagger X_{\bar\gamma}^\dagger|\phi\right\rangle \\
& +\sum_{\mu}\psi_i^{(5)}(\mu\gamma\gamma)\frac{1}{\sqrt{2}}
\left.a_\mu^\dagger (X_\gamma^\dagger)^2|\phi\right\rangle
  +\sum_{\mu}\psi_i^{(5)}(\mu\bar\gamma\bar\gamma)\frac{1}{\sqrt{2}}
\left.a_\mu^\dagger (X_{\bar\gamma}^\dagger)^2|\phi\right\rangle \\
& +\sum_{\bar\mu}\psi_i^{(5)}(\bar\mu\gamma\bar\gamma)
\left.a_{\bar\mu}^\dagger X_\gamma^\dagger X_{\bar\gamma}^\dagger|\phi\right\rangle \\
& +\sum_{\mu}\psi_i^{(7)}(\mu\gamma\gamma\gamma)\frac{1}{\sqrt{3!}}
\left.a_\mu^\dagger (X_\gamma^\dagger)^3|\phi\right\rangle
  +\sum_{\bar\mu}\psi_i^{(7)}(\bar\mu\bar\gamma\bar\gamma\bar\gamma)\frac{1}{\sqrt{3!}}
\left.a_{\bar\mu}^\dagger (X_{\bar\gamma}^\dagger)^3|\phi\right\rangle \\
& +\sum_{\bar\mu}\psi_i^{(7)}(\bar\mu\gamma\gamma\bar\gamma)\frac{1}{\sqrt{2}}
\left.a_{\bar\mu}^\dagger (X_\gamma^\dagger)^2 X_{\bar\gamma}^\dagger|\phi\right\rangle
  +\sum_\mu\psi_i^{(7)}(\mu\gamma\bar\gamma\bar\gamma)\frac{1}{\sqrt{2}}
\left.a_\mu^\dagger X_{\bar\gamma}^\dagger (X_{\bar\gamma}^\dagger)^2|\phi\right\rangle \\
& +\sum_{\mu}\psi_i^{(9)}(\mu\gamma\gamma\gamma\gamma)\frac{1}{\sqrt{4!}}
\left.a_\mu^\dagger (X_\gamma^\dagger)^4|\phi\right\rangle
  +\sum_\mu\psi_i^{(9)}(\mu\bar\gamma\bar\gamma\bar\gamma\bar\gamma)\frac{1}{\sqrt{4!}}
\left.a_\mu^\dagger (X_{\bar\gamma}^\dagger)^4|\phi\right\rangle \\
& +\sum_{\bar\mu}\psi_i^{(9)}(\bar\mu\gamma\gamma\gamma\bar\gamma)\frac{1}{\sqrt{3!}}
\left.a_{\bar\mu}^\dagger (X_\gamma^\dagger)^3 X_{\bar\gamma}^\dagger|\phi\right\rangle
  +\sum_{\bar\mu}\psi_i^{(9)}(\bar\mu\gamma\bar\gamma\bar\gamma\bar\gamma)\frac{1}{\sqrt{3!}}
\left.a_{\bar\mu}^\dagger X_\gamma^\dagger (X_{\bar\gamma}^\dagger)^3|\phi\right\rangle \\
& +\sum_\mu\psi_i^{(9)}(\mu\gamma\gamma\bar\gamma\bar\gamma)\frac{1}{2}
\left.a_\mu^\dagger (X_\gamma^\dagger)^2 (X_{\bar\gamma}^\dagger)^2|\phi\right\rangle , \\
&\mbox{for the $r=-i$ sector} ,
\label{wf}
\end{split}
\end{equation}
where $\gamma$ and $\bar\gamma$ abbreviate $\gamma(+)$ and $\gamma(-)$, 
respectively, and $\left.|\phi\right\rangle$ is the rotating vacuum 
configuration. Those for the $r=+i$ sector take a similar form. 
This notation indicates that a limited class of 1, 3, 5, 7, and 9qp states that contribute 
to multi-phonon $\gamma$-vibrational states are taken into account. Among these, 
the model space was truncated up to the $\psi^{(5)}$ terms in Ref.~\cite{MM1}. 
Here it should be noted that the tilt of the rotational axis brought about by large 
triaxial deformations can mix signature quantum number in general. But the rotation is 
essentially one-dimensional when the static triaxial deformation is smaller than the zero-point 
amplitude of the $\gamma$ vibration. 

\subsection{Generalized intensity relation}

The rotational effects on the vibrational (interband) transition rates are well described 
by the generalized intensity relation in terms of the intrinsic matrix elements, see 
Fig.~4-30 in Ref.~\cite{BM}. 
On the other hand, the cranking model and its extensions can provide us 
with rotationally perturbed intrinsic matrix elements precisely. Therefore, a method to 
combine these two was proposed in Ref.~\cite{SN} and applied to the 
$1\gamma\rightarrow 0\gamma$ transitions in 
$^{165}$Ho and $^{167}$Er in Ref.~\cite{Ge}. In the present study, this method is applied to the 
$n\gamma\rightarrow(n-1)\gamma$ transitions with $n=$1, 2, and 3, along 
the way of this reference. The expressions for the $B(E2)$ are 
\begin{gather}
B(E2:I_\mathrm{i}K_\mathrm{i}\rightarrow I_\mathrm{f}K_\mathrm{f})=\langle I_\mathrm{i}K_\mathrm{i}2\Delta K|I_\mathrm{f}K_\mathrm{f}\rangle^2 Q_\mathrm{out}^2, \\
Q_\mathrm{out}=Q_1+Q_2[I_\mathrm{f}(I_\mathrm{f}+1)-I_\mathrm{i}(I_\mathrm{i}+1)], \\
Q_1=\sqrt{2}Q_\mathrm{tr}-\Delta K(K_\mathrm{i}+K_\mathrm{f})Q_2, \\
Q_\mathrm{tr}=\langle\mathrm{f}|Q_2^{(+)}|\mathrm{i}\rangle, \\
Q_2=\frac{1}{\sqrt{2}\mathcal{J}}\frac{\mathrm{d}\langle\mathrm{f}|Q_1^{(+)}|\mathrm{i}\rangle}{\mathrm{d}\omega_\mathrm{rot}}, 
\end{gather}
where $Q_\mathrm{tr}$ and $Q_2$ are evaluated at $\omega_\mathrm{rot}=0$, and the moment 
of inertia, $\mathcal{J}$, is extracted from the experimental energy of the first $\Delta I=2$ 
excited states in the ground band. 

\section{Results and discussion}

Numerical calculations are performed for two isotopes, $^{103}$Nb and $^{105}$Nb, in which 
candidates of $3\gamma$ states were observed~\cite{Nb103,Nb105}. The former was investigated 
in Ref.~\cite{MM1} within a limited model space, and the latter, on which the data were reported quite 
recently, is newly studied. In both isotopes, the ground band is based on the $\pi[422]\,5/2^+$ 
asymptotic state, and the single- and multi-phonon $\gamma$-vibrational excitations on top of 
it were observed. Cranking and RPA calculations are done in the five major shells, 
$N_\mathrm{osc}=2$ -- 6 for the neutron and 1 - 5 for the proton. The indices $\mu$ and 
$\bar\mu$ in Eq.~(\ref{wf}) for the PVC eigenstates run from 1 to 15, the number of quasiparticle 
states with $N_\mathrm{osc}=4$. 
In the following, the results for the favored signature, $r=-i$, are 
mainly presented while those for the unfavored $r=+i$ are also shown when necessary. 

\subsection{$0\gamma$ -- $2\gamma$ states in $^{103}$Nb}

All the parameters entering into the calculation are the same as those adopted in Ref.~\cite{MM1}. 
The pairing gaps $\Delta_n=1.05$ MeV, $\Delta_p=0.85$ MeV and the deformation 
$\epsilon_2=0.31$ are adopted conforming to the experimental analyses~\cite{Mo106,Nb103}. 
The triaxiality $\gamma=-7^\circ$ is chosen to reproduce the measured signature splitting 
of the ground band in the PVC calculation. The quadrupole force strengths are determined 
to reproduce in the axially symmetric limit the $\gamma$-vibrational energy observed in 
$^{104}$Mo~\cite{Mo104}, see Ref.~\cite{MM1} for the detail. 

\subsubsection{Distribution of collective states}

\begin{figure}[htbp]
 \includegraphics[width=6cm]{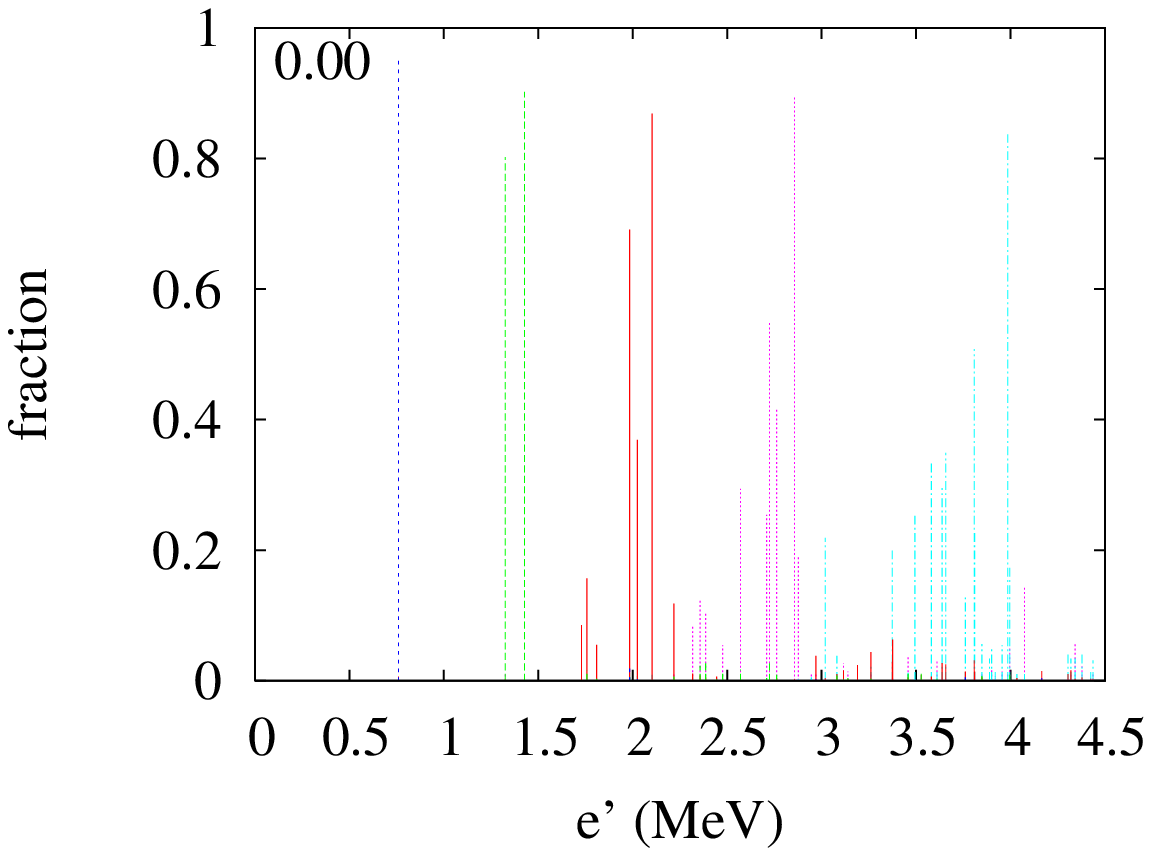}
 \includegraphics[width=6cm]{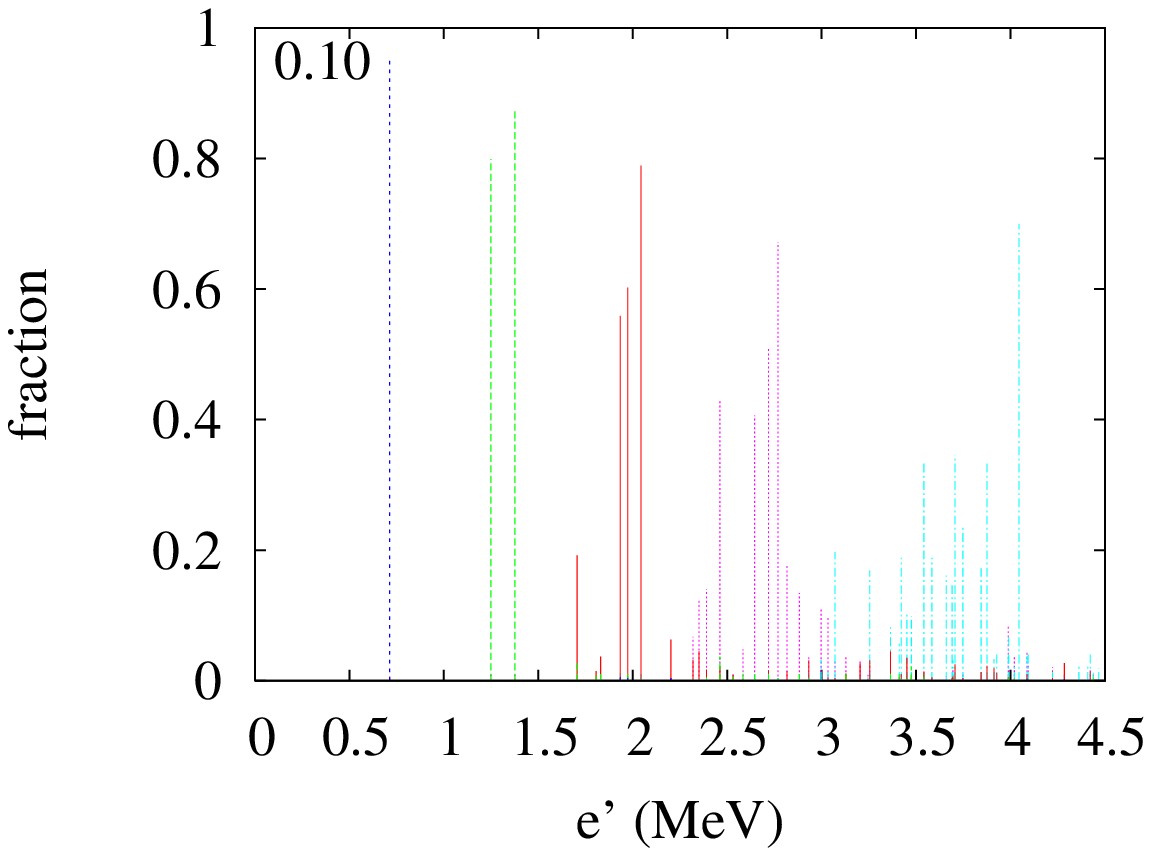}
 \includegraphics[width=6cm]{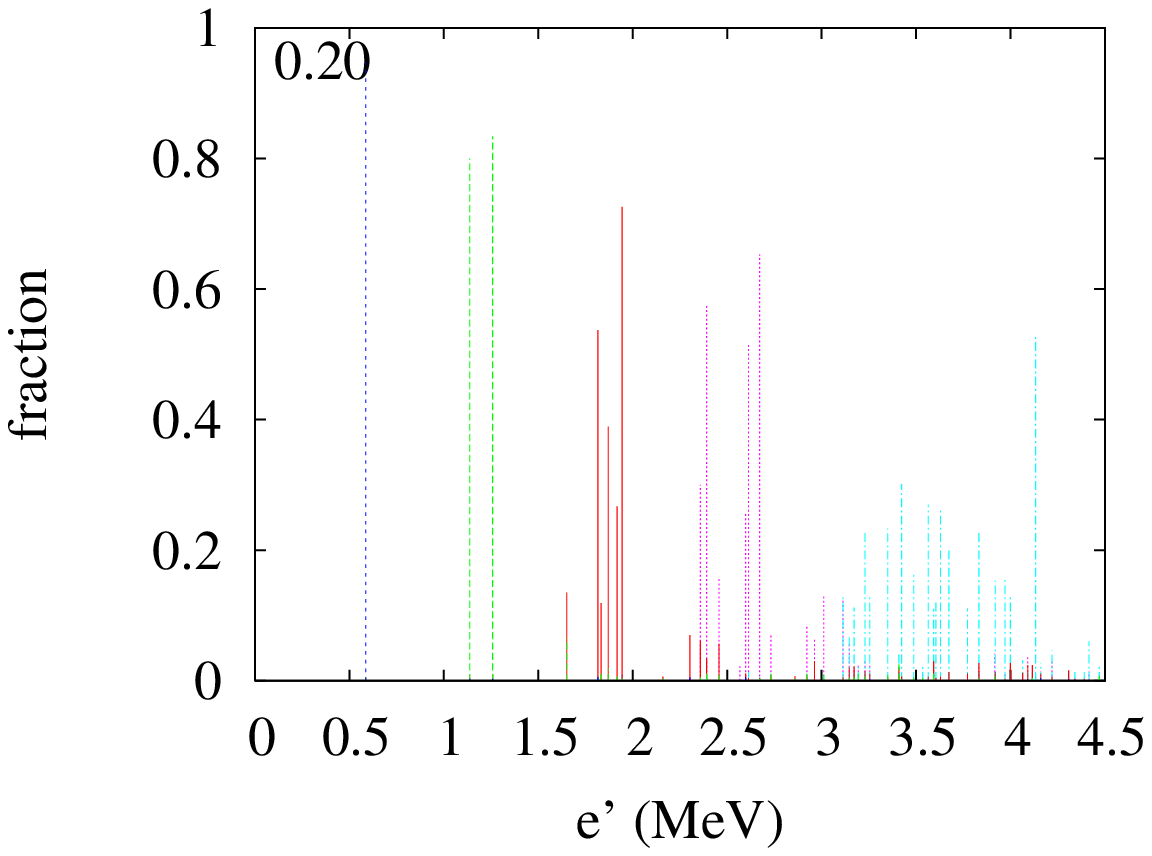}
 \includegraphics[width=6cm]{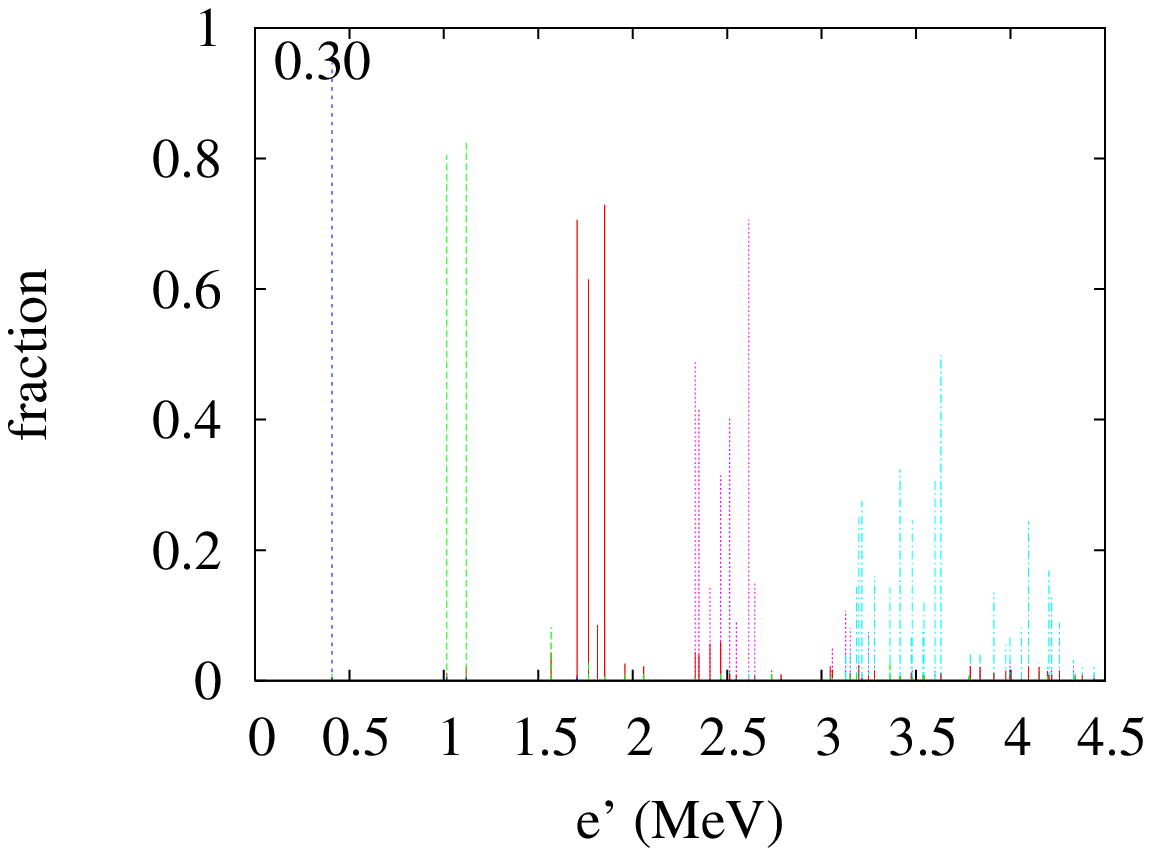}
 \caption{(Color online) Distribution of the collective fraction (probability in the 
wave function) of the $0\gamma$, $1\gamma$, $2\gamma$, $3\gamma$, and $4\gamma$ components 
in the favored signature ($r=-i$) sector of $^{103}$Nb, 
$|\psi^{(1)}(1)|^2$ (blue longer dotted), 
$|\psi^{(3)}(1\gamma)|^2+|\psi^{(3)}(\bar 1 \bar\gamma)|^2$ (green dashed), 
$|\psi^{(5)}(1\gamma\gamma)|^2+|\psi^{(5)}(1\bar\gamma\bar\gamma)|^2+
 |\psi^{(5)}(\bar 1 \gamma\bar\gamma)|^2$ (red solid), 
$|\psi^{(7)}(1\gamma\gamma\gamma)|^2+|\psi^{(7)}(\bar 1 \bar\gamma\bar\gamma\bar\gamma)|^2+
 |\psi^{(7)}(\bar 1 \gamma\gamma\bar\gamma)|^2+|\psi^{(7)}(1\gamma\bar\gamma\bar\gamma)|^2$ 
(magenta dotted), and 
$|\psi^{(9)}(1\gamma\gamma\gamma\gamma)|^2+|\psi^{(9)}(1\bar\gamma\bar\gamma\bar\gamma\bar\gamma)|^2+
 |\psi^{(9)}(\bar 1 \gamma\gamma\gamma\bar\gamma)|^2+
 |\psi^{(9)}(\bar 1 \gamma\bar\gamma\bar\gamma\bar\gamma)|^2+
 |\psi^{(9)}(1\gamma\gamma\bar\gamma\bar\gamma)|^2$ (blue dot-dashed), respectively, 
at $\omega_\mathrm{rot}=0$ -- 0.3 MeV.}
 \label{fig1}
\end{figure}

In Fig.~\ref{fig1}, the distribution of collective states are shown. In the following, 
I denote the favored signature of the lowest quasiparticle state as $1qp$, and its 
signature partner as $\overline{1qp}$. The lowest PVC eigenstate whose main component is 
this $1qp$ is often denoted also as the $0\gamma$ state. Then the fraction of the $1\gamma$ 
components (green dashed in the figure) means the sum of the probabilities of $1qp\otimes\gamma(+)$ 
and $\overline{1qp}\otimes\gamma(-)$ basis states. The conventions for multi-phonon states are 
understood straightforwardly. 

At $\omega_\mathrm{rot}=0$, $0\gamma$ -- $3\gamma$ states are almost harmonic, whereas the 
$4\gamma$ strength is located higher than the harmonic location. In addition, 
the main peaks of $0\gamma$ -- $4\gamma$ states are almost of the same height although the 
collective strength spreads as the number of phonon increases. As discussed later, 
the main peaks have fairly pure $K$. 

As soon as rotation sets in, the heights of the main peaks decrease approximately 
in proportion to the number of phonon. Two sequences of $1\gamma$ and three 
sequences of $2\gamma$ states survive up to high spins as discussed in Ref.~\cite{MM1}. 
In addition, three or four sequences of $3\gamma$ states keep their collective 
character to some extent. 
Routhians of $n\gamma$ states are $e'_{2\gamma}<e'_{0\gamma}+2\Delta_p$, 
but $e'_{0\gamma}+2\Delta_p<e'_{3\gamma}<e'_{0\gamma}+2\Delta_n$. 
Therefore, the collectivity of the calculated $3\gamma$ states other than high-$K$ 
ones, which are hard to mix with other states, should be considered with reservations, 
because in the present model $1qp\otimes 3\gamma$ basis states couple only with 
(1qp)$'\otimes 2\gamma$ and (1qp)$'\otimes 4\gamma$, whereas direct couplings to 3qp 
states are not included, where (1qp)$'$ denotes other one-quasiparticle basis states including 
those of the opposite signature. 

\begin{figure}[htbp]
 \includegraphics[width=6cm]{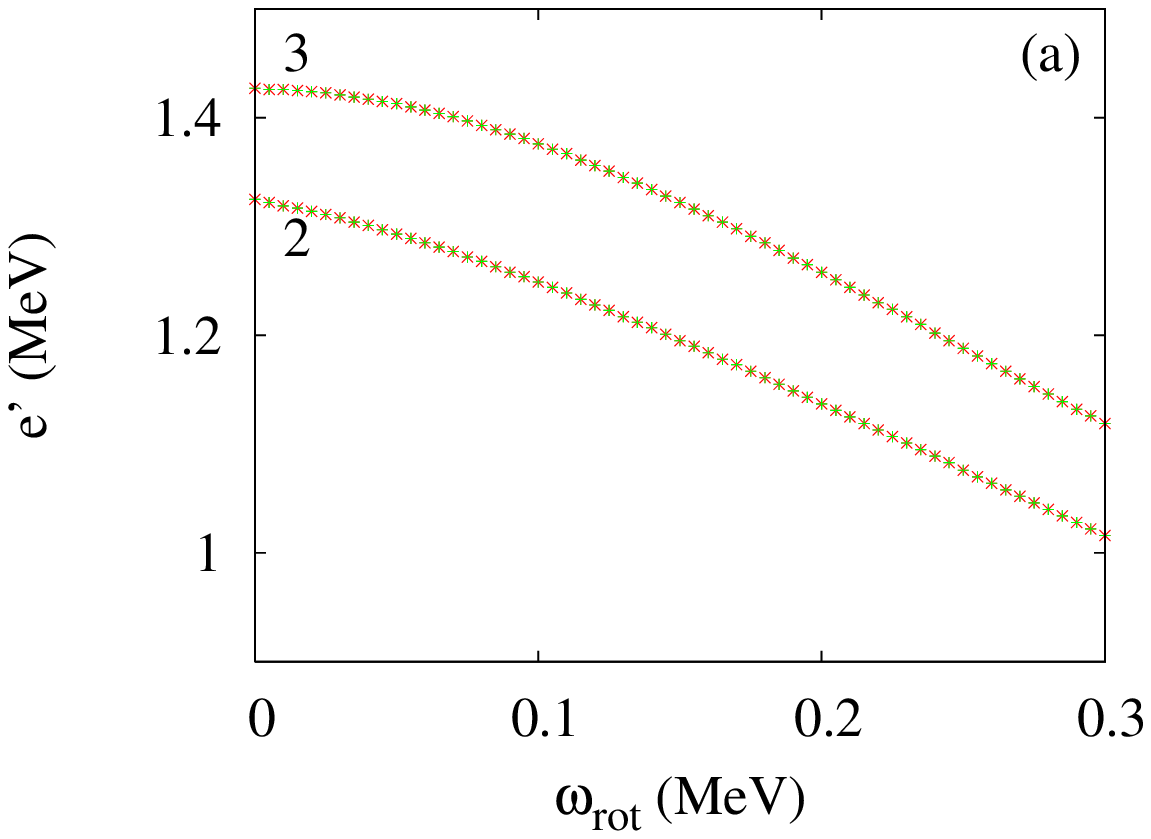}
 \includegraphics[width=6cm]{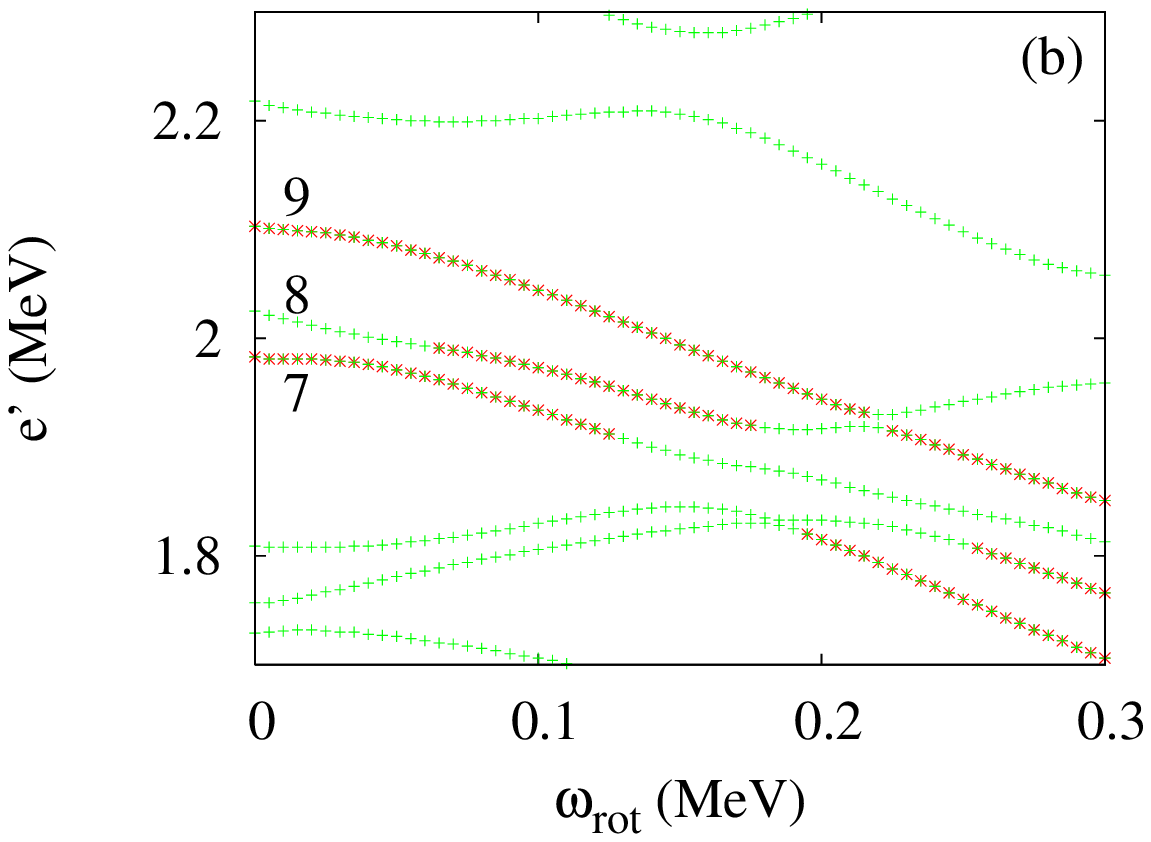}
 \caption{(Color online) Routhians of all calculated PVC states in the $r=-i$ sector 
of $^{103}$Nb 
in the regions of (a) $1\gamma$ and (b) $2\gamma$ bands are shown by green $+$s. 
Those with more than 50\% collective fraction are emphasized by red $\times$s. 
The labels attached designate the numbers, $i$ in Eq.~(\ref{wf}), enumerated from the 
lowest.}
 \label{fig2}
\end{figure}

In Fig.~\ref{fig2}, calculated eigenstates in the regions of (a) $1\gamma$ 
and (b) $2\gamma$ states are shown. Among them, those with more than 50\% collective 
($1\gamma$ or $2\gamma$) fraction are emphasized with red crosses. The two $1\gamma$ 
sequences are completely isolated from other states. The three $2\gamma$ ones are 
also distinguished from other states but crossed by two up-slope noncollective states at 
around $\omega_\mathrm{rot}=0.15$ -- 0.2 MeV. 

\subsubsection{Characterization of the calculated $1\gamma$ and $2\gamma$ states}

\begin{figure}[htbp]
 \includegraphics[width=6cm]{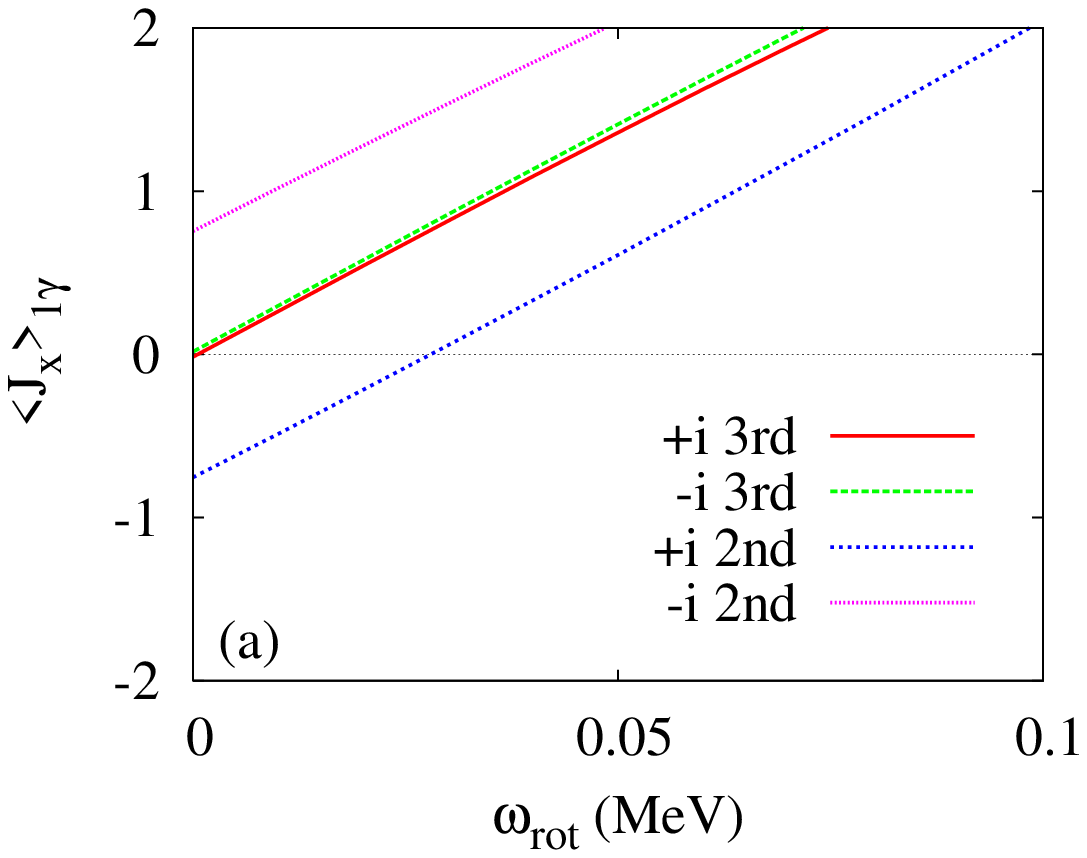}
 \includegraphics[width=6cm]{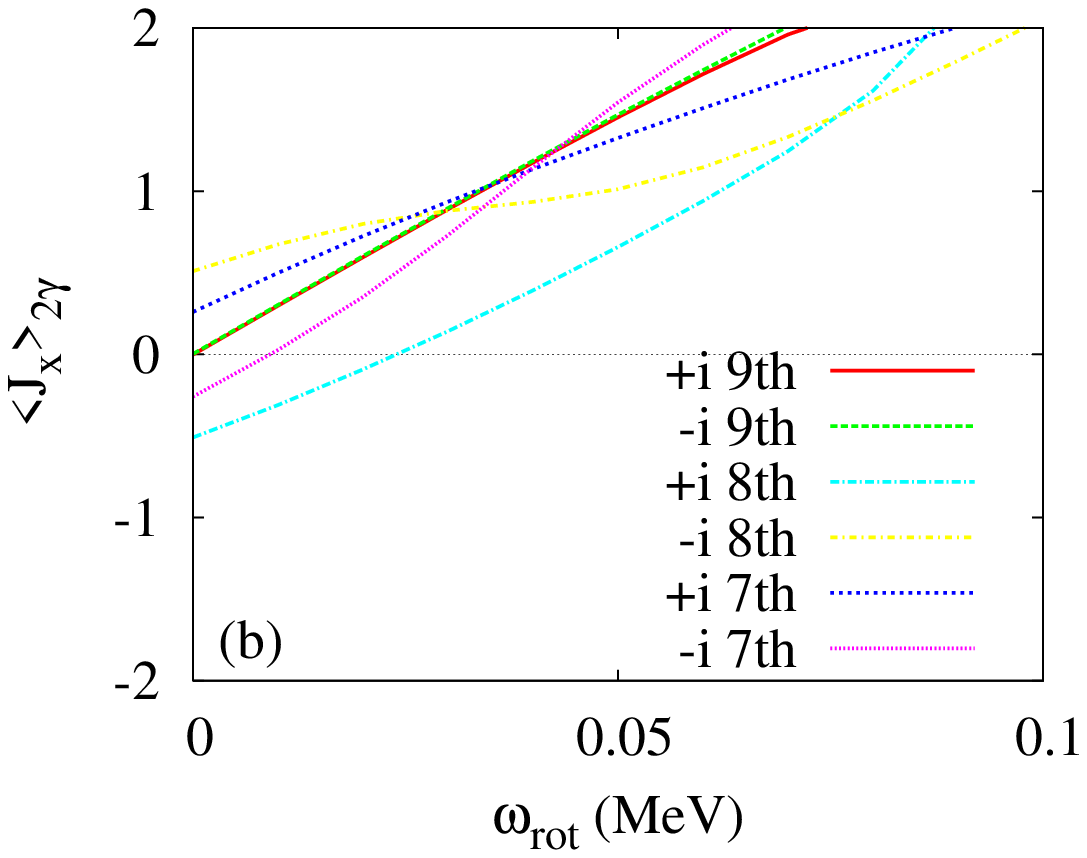}
 \caption{(Color online) Expectation values of the projection of the angular momentum 
to the rotational ($x$) axis around the band heads of (a) $1\gamma$ and (b) $2\gamma$ 
bands in $^{103}$Nb.}
 \label{fig3}
\end{figure}

On the correspondence between $1\gamma$ bands in the calculation in the signature 
scheme and that in the $K$ scheme, it was argued in Ref.~\cite{MM1} that the obtained lower band 
can be identified with the $K=\Omega-2$ band because states with lower $K$ have lower 
intrinsic energies than those with higher $K$ and the same total angular momentum $I$. 
To look into this correspondence more, the aligned angular momenta around 
$\omega_\mathrm{rot}=0$ are shown in Figs.~\ref{fig3} (a) for $1\gamma$ and (b) for 
$2\gamma$ states. The labels $i$ of states refer to those in Fig.~\ref{fig2}. 

The $1\gamma$ band should have $K=|\Omega-2|=1/2$ or $\Omega+2=9/2$ at the band head 
aside from weak $K$ mixing stemming from static triaxial deformation. 
Figure~\ref{fig3}(a) clearly shows that the lower (second) pair of states has a strong 
signature decoupling, non-zero aligned angular momentum with opposite sign at 
$\omega_\mathrm{rot}=0$, while the upper (third) pair has practically zero 
aligned angular momentum at $\omega_\mathrm{rot}=0$ and negligible signature splitting. 
This proves the mapping that the lower band is of low $K$ and the upper band is of 
high $K$. 

\begin{figure}[htbp]
 \includegraphics[width=6cm]{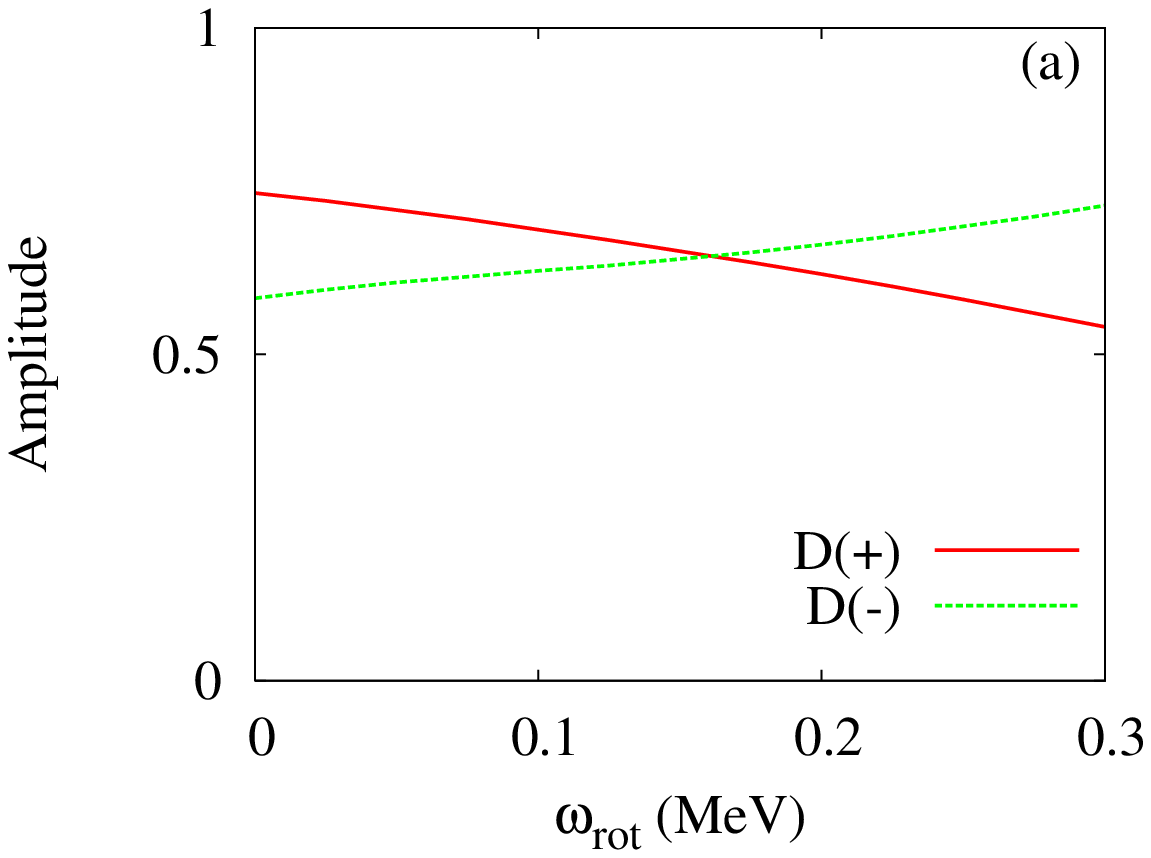}
 \includegraphics[width=6cm]{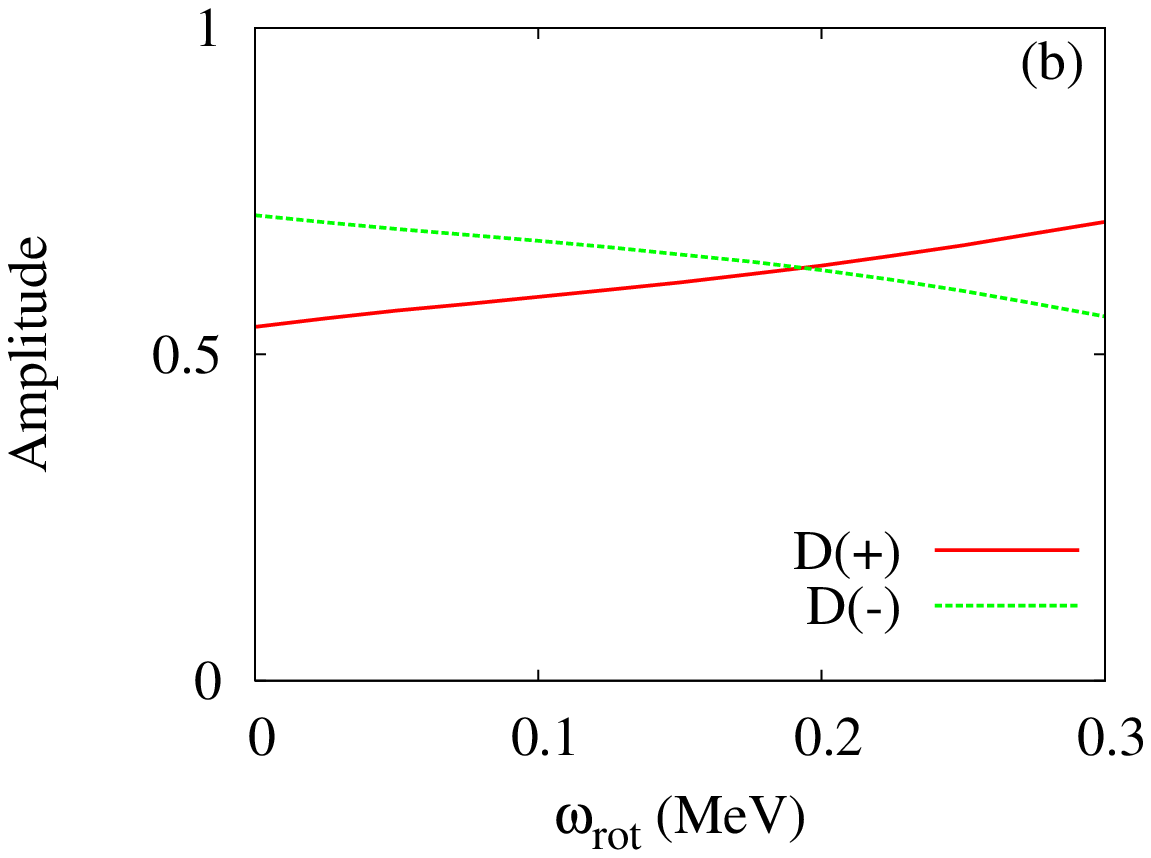}
 \caption{(Color online) Amplitudes of the dominant components 
$D(+)=|\psi_i^{(3)}(1\gamma)|$ and $D(-)=|\psi_i^{(3)}(\bar 1 \bar\gamma)|$ 
in (a) upper ($i=$ 3rd) and (b) lower ($i=$ 2nd) 
$1\gamma$ bands in the $r=-i$ sector of $^{103}$Nb.}
 \label{fig4}
\end{figure}

The characterization is further evidenced by looking at the wave function. 
Figure~\ref{fig4} graphs the amplitudes of the dominant components, 
$D(+)=|\psi_i^{(3)}(1\gamma)|$ for $1qp\otimes\gamma(+)$ and 
$D(-)=|\psi_i^{(3)}(\bar 1 \bar\gamma)|$ for $\overline{1qp}\otimes\gamma(-)$, 
in (a) upper ($i=$ 3rd) and (b) lower ($i=$ 2nd) bands. The structure of the 
$\gamma(\pm)$ is $(Q_{22}\pm Q_{2-2})/\sqrt{2}$. These two components 
mix with similar magnitudes both in the upper and lower eigenstates 
and their relative sign (not shown) is always opposite. This means that the 
two orthogonal combinations of $\gamma(\pm)$ reproduce high-$K$ and low-$K$ 
states. 

The $2\gamma$ band should have $K=|\Omega-4|=3/2$ or $\Omega=5/2$ or 
$\Omega+4=13/2$ at the band head. An argument for Fig.~\ref{fig3}(b) similar 
to that for $1\gamma$ bands in Fig.~\ref{fig3}(a) leads to the identification 
that the ninth pair is $K=13/2$, the eighth and seventh are $K=3/2$ and $K=5/2$. The latter 
two interact with each other in $r=-i$ as soon as rotation sets in. This 
interaction can also be seen in Fig.~\ref{fig2}(b). Note that the eighth state 
has about 35\% collectivity at $\omega_\mathrm{rot}=0$ but it increases 
as $\omega_\mathrm{rot}$ increases. 

This discussion confirms that the calculated upper bands, third for $1\gamma$ 
and ninth for $2\gamma$, possess the character of the highest $K$, $K=9/2$ and 13/2, 
respectively. Because the observed bands (2) and (3) were assigned experimentally 
as $K=9/2$ and 13/2, respectively, the correspondence between the theory and 
experiment is established. This is natural in that the most collective state 
with less mixings with noncollective states would be observed. 

\subsubsection{Effect of enlargement of the model space and comparison with 
experimental data}

The first purpose of this paper is to see how the $2\gamma$ states calculated 
in Ref.~\cite{MM1} within a smaller model space are affected by the enlargement of 
the space. The previous calculation was done in the space up to $2\gamma$ basis states. 
This time I examine that up to $3\gamma$ and $4\gamma$ basis states. 
First, by including $3\gamma$ basis states, the upper $2\gamma$ band is pushed down by 
0.27 MeV ($\omega_\mathrm{rot}=0$) -- 0.23 MeV ($\omega_\mathrm{rot}=0.3$ MeV). 
Next, by including $4\gamma$, this band is pushed down further by 0.06 -- 0.03 
MeV. The calculated $0\gamma$ and $1\gamma$ states are almost unaffected. 

\begin{figure}[htbp]
 \includegraphics[width=6cm]{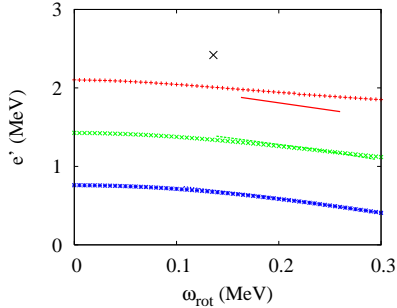}
 \caption{(Color online) Routhians of calculated  $0\gamma$ (blue $\ast$s), $1\gamma$ 
(green $\times$s), and $2\gamma$ (red $+$s) states in the $r=-i$ sector of $^{103}$Nb are 
compared with the corresponding data (curves) converted to the rotating frame 
by using the Harris parameters $\mathcal{J}_0=15.45$ MeV$^{-1}$ and 
$\mathcal{J}_1=81.23$ MeV$^{-3}$ that fit the yrast band of $^{104}$Mo~\cite{Mo104}. 
The observed transition in band (4), the $3\gamma$ candidate, converted to the rotating 
frame is also shown by a large $\times$.}
 \label{fig5}
\end{figure}

The final result is presented in Fig.~\ref{fig5}. In this nucleus, calculated 
$e'_{2\gamma}$ is still higher, by 0.09 -- 0.18 MeV, than the data. See the 
result for $^{105}$Nb below. 

\subsection{$0\gamma$ -- $2\gamma$ states in $^{105}$Nb}

\begin{figure}[htbp]
 \includegraphics[width=6cm]{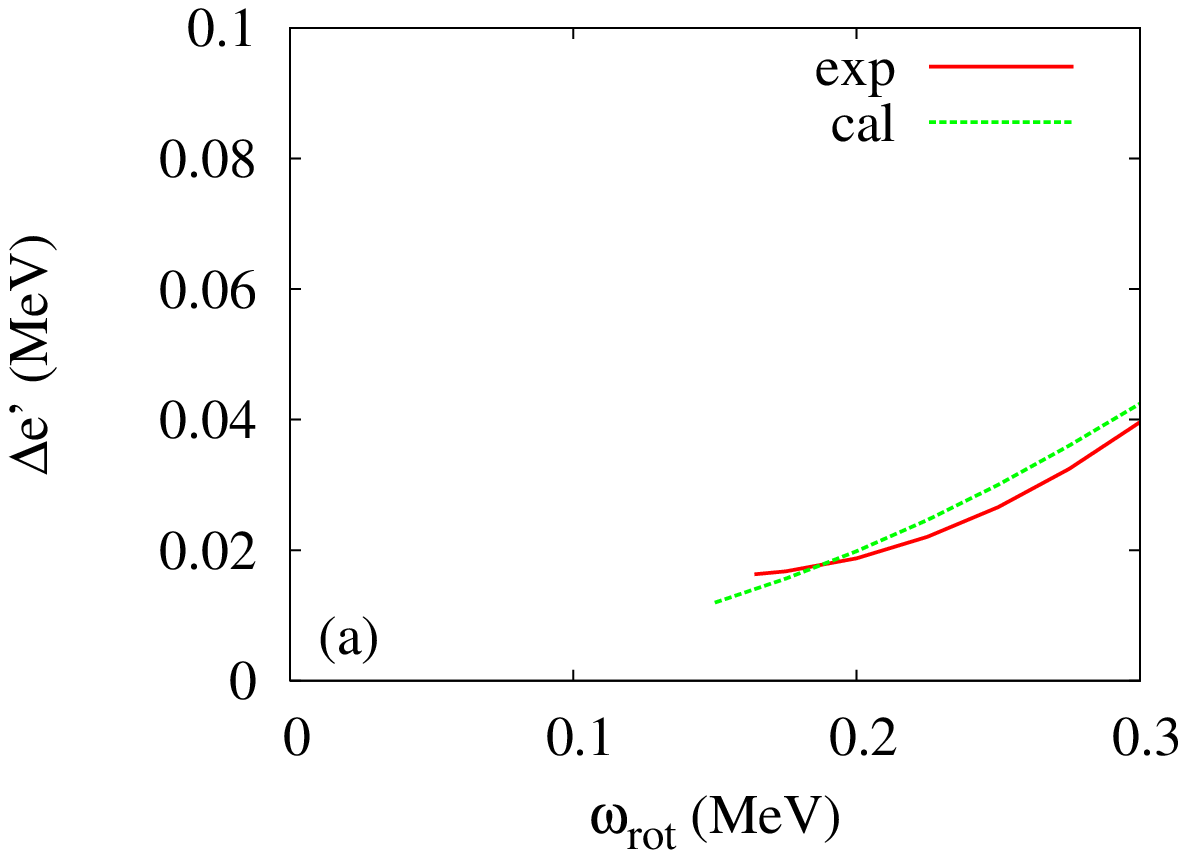}
 \includegraphics[width=6cm]{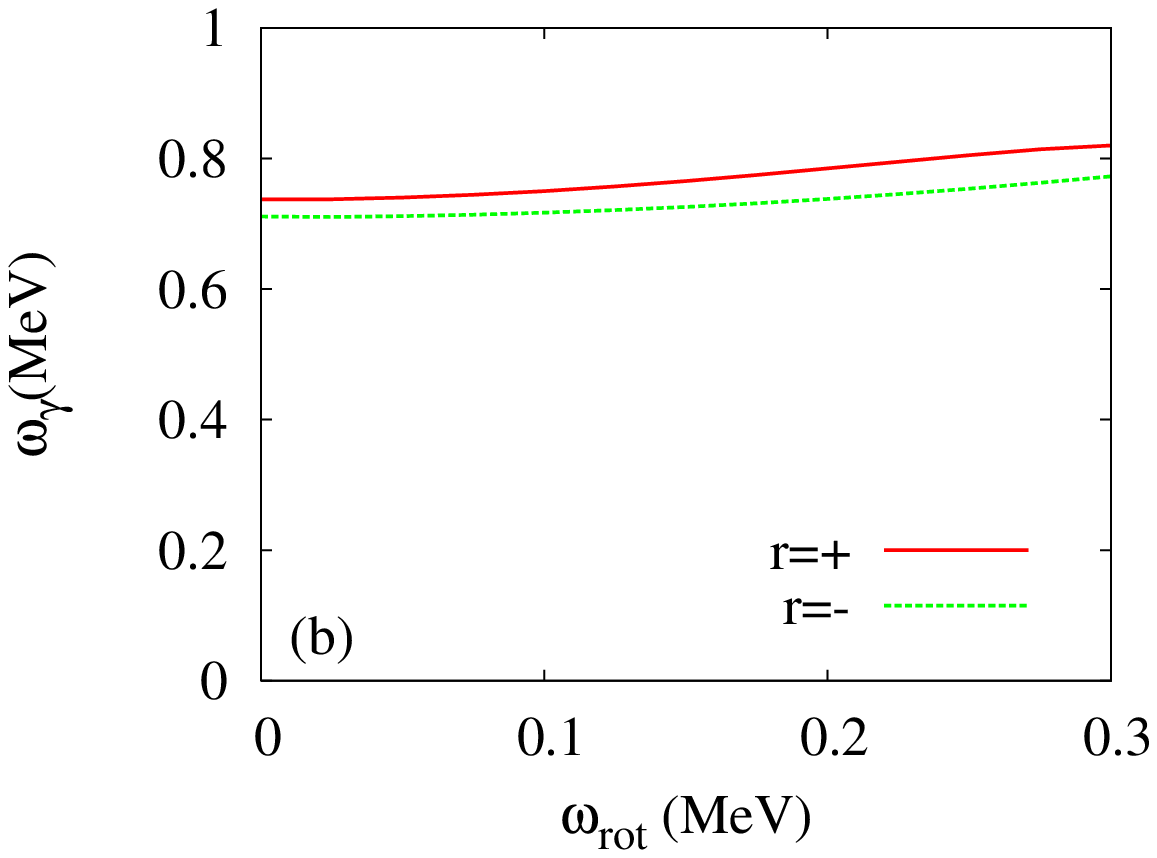}
 \caption{(Color online) (a) Experimental and calculated signature splitting 
in the $\pi[422]\,5/2^+$ one-quasiparticle band in $^{105}$Nb. 
Theoretical curve is the result of the particle-vibration coupling calculation. 
(b) Excitation energies of $\gamma$-vibrational RPA phonons in the rotating frame with $r=\pm1$.}
 \label{fig6}
\end{figure}

Parameters entering into the calculation are determined in a manner similar to 
the case of $^{103}$Nb. Concretely, the pairing gaps $\Delta_n=1.05$ MeV and 
$\Delta_p=0.85$ MeV, and the deformation $\epsilon_2=0.3254$ are 
adopted conforming to the experimental analyses~\cite{Mo106,Nb105}. 
The triaxiality $\gamma=-10^\circ$ is chosen to reproduce the observed signature 
splitting of the ground ($0\gamma$) band in the PVC calculation as shown in Fig.~\ref{fig6}(a). 

The way to determine the quadrupole force strengths is slightly different; 
those determined to reproduce $\omega_\gamma=0.7104$ MeV of $^{106}$Mo~\cite{Mo106} 
in the reference configuration with $\omega_\mathrm{rot}=0$ and $\gamma=0$ result in 
a large signature splitting in $\omega_\gamma$ when triaxial deformation is 
introduced in contrast to the case of $^{103}$Nb. Alternatively, they are adjusted 
so as to reproduce the above $\omega_\gamma$ at $\omega_\mathrm{rot}=0$ and 
$\gamma=-10^\circ$, then $\kappa_0^{(+)}$ is set equal to $\kappa_2^{(+)}$ as in the 
case of $^{103}$Nb. The values for the residual pairing interaction are set to 
reproduce the adopted pairing gaps. The obtained $\omega_\mathrm{rot}$ dependence and the 
signature splitting of $\omega_\gamma$ are shown in Fig.~\ref{fig6}(b). 

\subsubsection{Distribution of collective states}

\begin{figure}[htbp]
 \includegraphics[width=6cm]{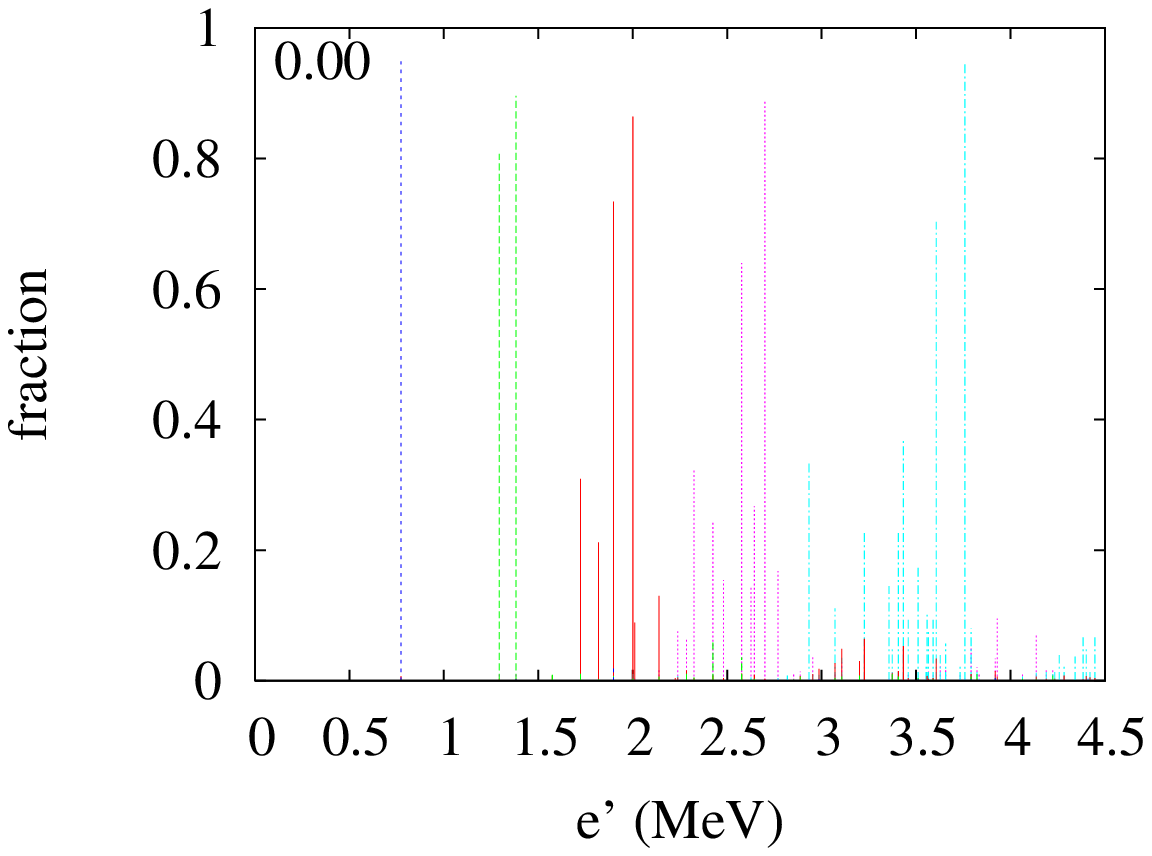}
 \includegraphics[width=6cm]{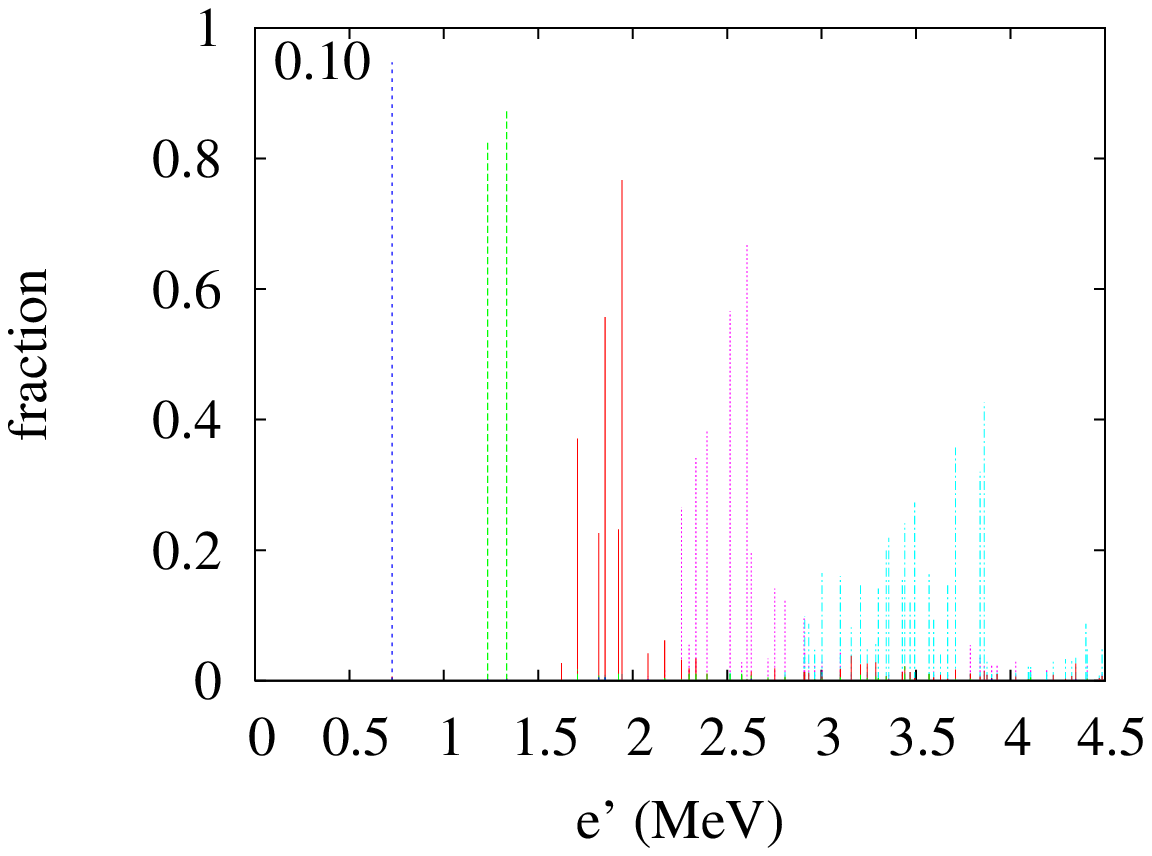}
 \includegraphics[width=6cm]{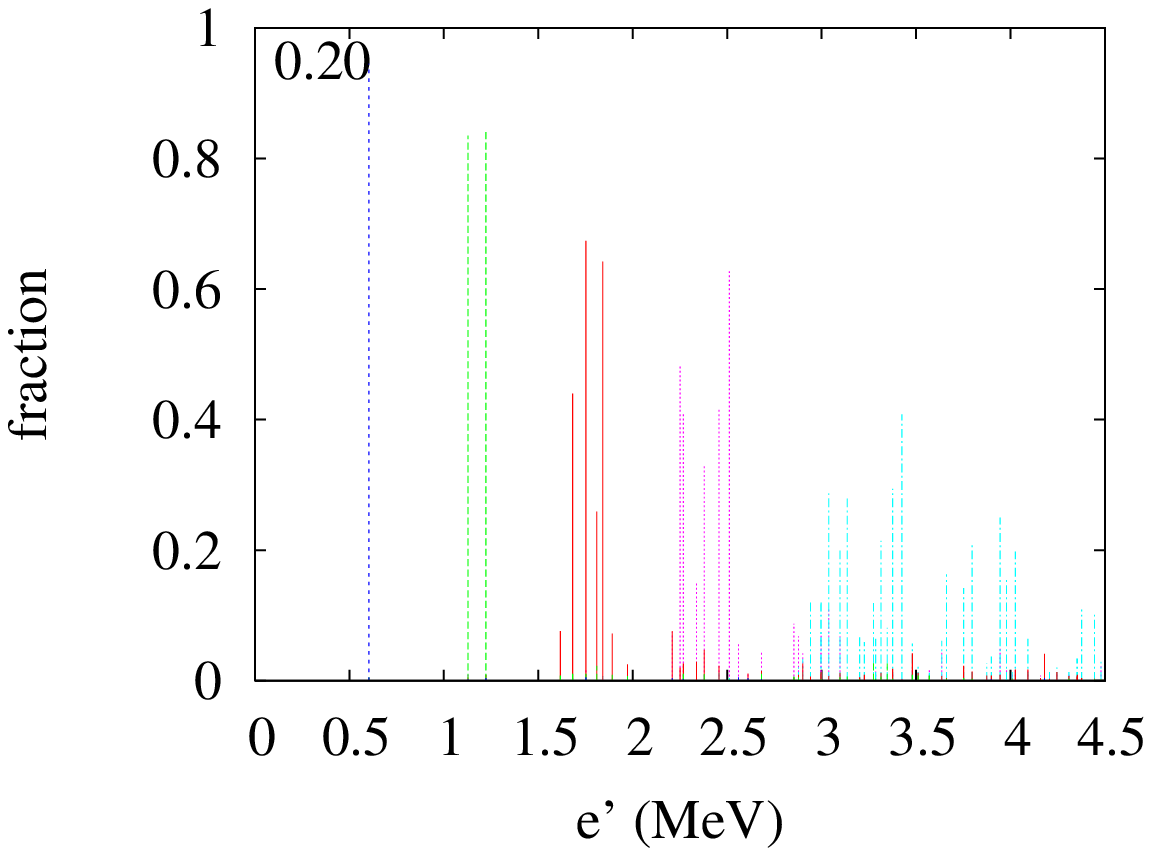}
 \includegraphics[width=6cm]{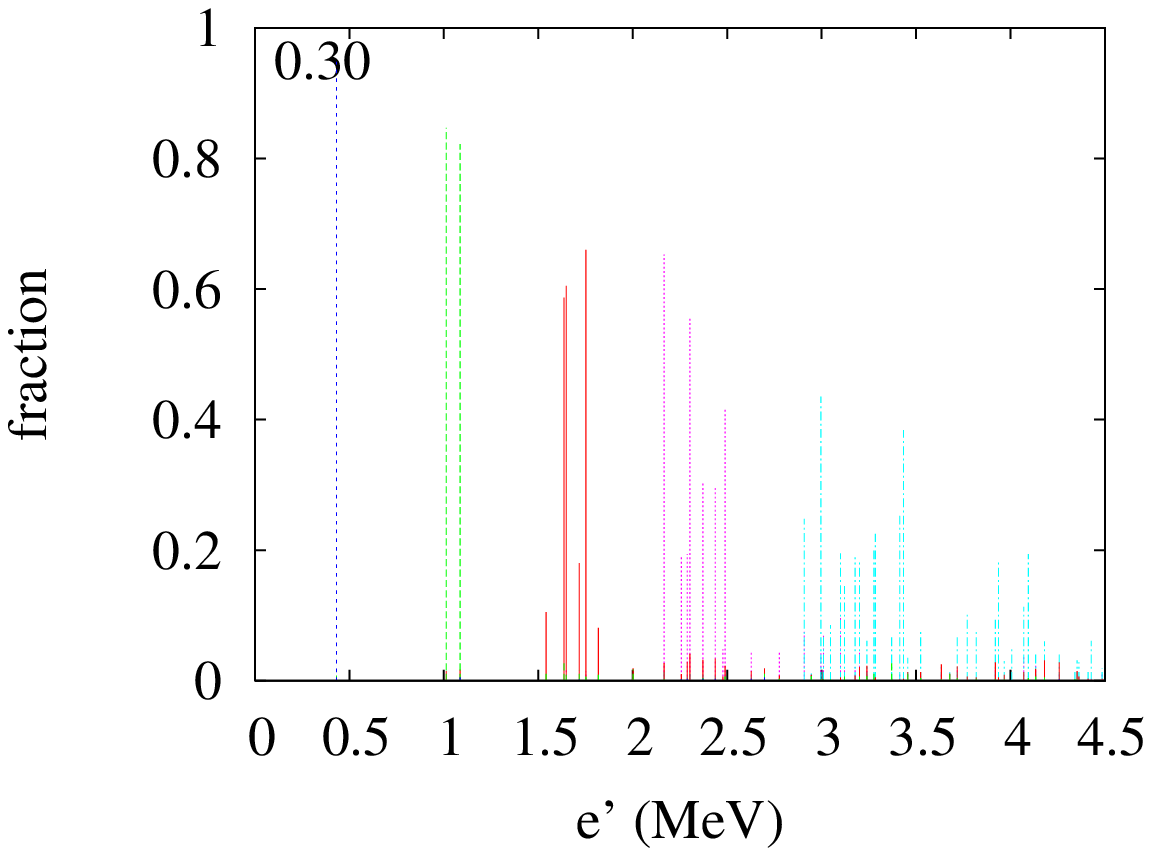}
 \caption{(Color online) The same as Fig.~\ref{fig1} but for $^{105}$Nb.}
 \label{fig7}
\end{figure}

\begin{figure}[htbp]
 \includegraphics[width=6cm]{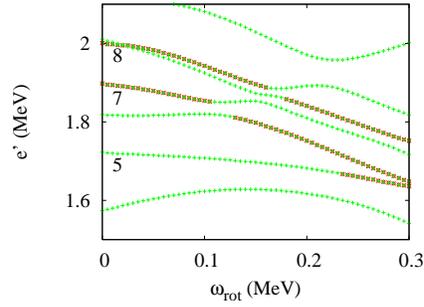}
 \caption{(Color online) The same as Fig.~\ref{fig2}(b) but for $^{105}$Nb.}
 \label{fig8}
\end{figure}

In Fig.~\ref{fig7}, the distribution of collective states is shown. Overall 
feature is quite similar to the case of $^{103}$Nb but the Routhians of collective 
solutions are slightly lower reflecting the input $\omega_\gamma$; this is consistent 
with the data, see Fig.~7 in Ref.~\cite{Nb105}. Other differences from the $^{103}$Nb 
case are that (i) the collectivity of the third strongest $2\gamma$ (fifth) state is as low 
as 30\% at around $\omega_\mathrm{rot}=0$ but increases as $\omega_\mathrm{rot}$ 
increases as seen in Fig.~\ref{fig8}, and (ii) among the $3\gamma$ states, the lower one 
is the most collective at $\omega_\mathrm{rot}=0.3$ MeV. The latter feature will 
be discussed below. 

\subsubsection{Comparison with experimental data}

\begin{figure}[htbp]
 \includegraphics[width=6cm]{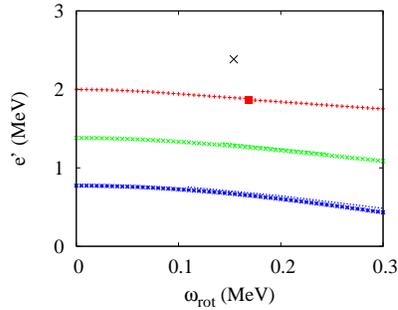}
 \caption{(Color online) The same as Fig.~\ref{fig5} but for $^{105}$Nb.
The Harris parameters $\mathcal{J}_0=$ 18.08 MeV$^{-1}$ and 
$\mathcal{J}_1=$ 43.21 MeV$^{-3}$ that fit the yrast band of $^{106}$Mo~\cite{Mo106} 
were used for the conversion.}
 \label{fig9}
\end{figure}

The characterization of calculated states is done in the same manner as in 
the case of $^{103}$Nb. Then the comparison with the data of 
$0\gamma$ -- $2\gamma$ states are shown in Fig.~\ref{fig9}. In the present 
case the observed $2\gamma$ state is perfectly reproduced in contrast to the $^{103}$Nb 
case in which some deviation remains. 

\subsection{$3\gamma$ states in $^{103}$Nb and $^{105}$Nb}

\subsubsection{Distribution of collective states}

An issue beyond the scope of Ref.~\cite{MM1} is to characterize the observed band (4) 
that is conjectured to be a candidate of the $3\gamma$ state, the first 
three-phonon state in deformed nuclei if confirmed. 

\begin{figure}[htbp]
 \includegraphics[width=6cm]{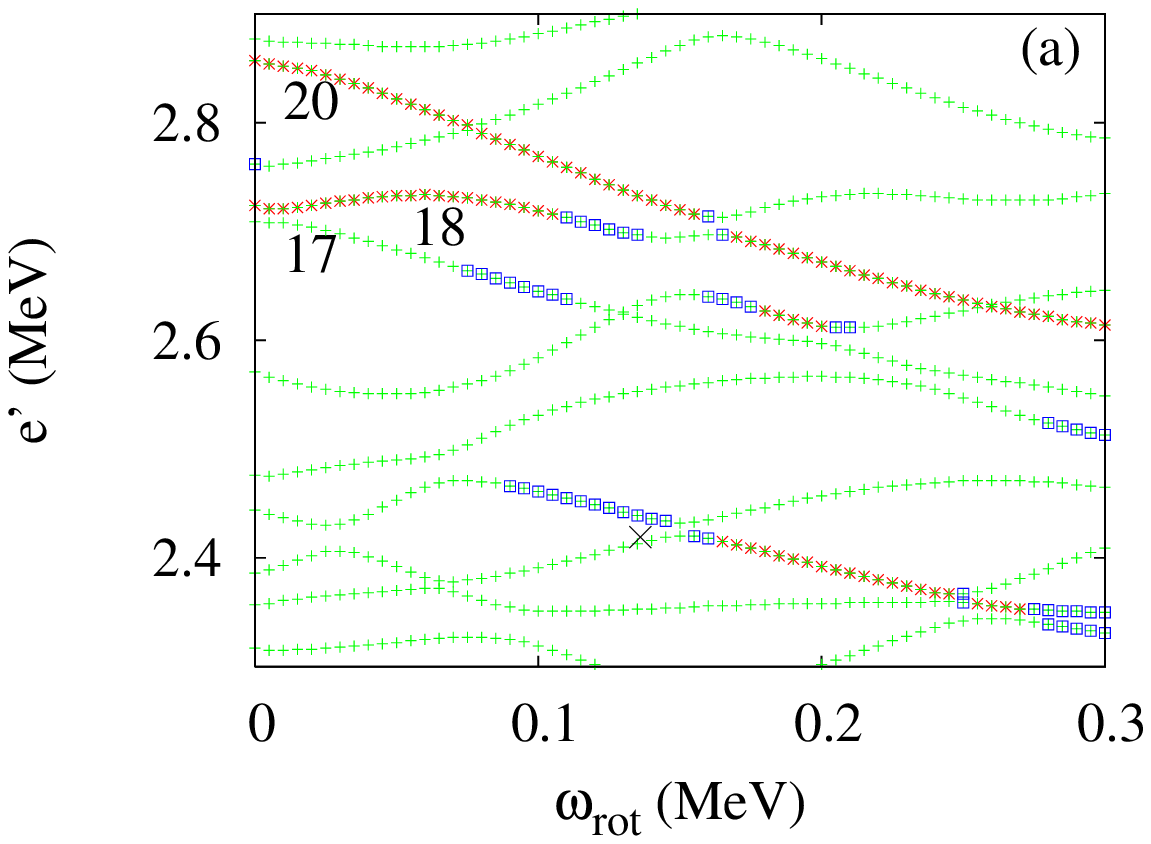}
 \includegraphics[width=6cm]{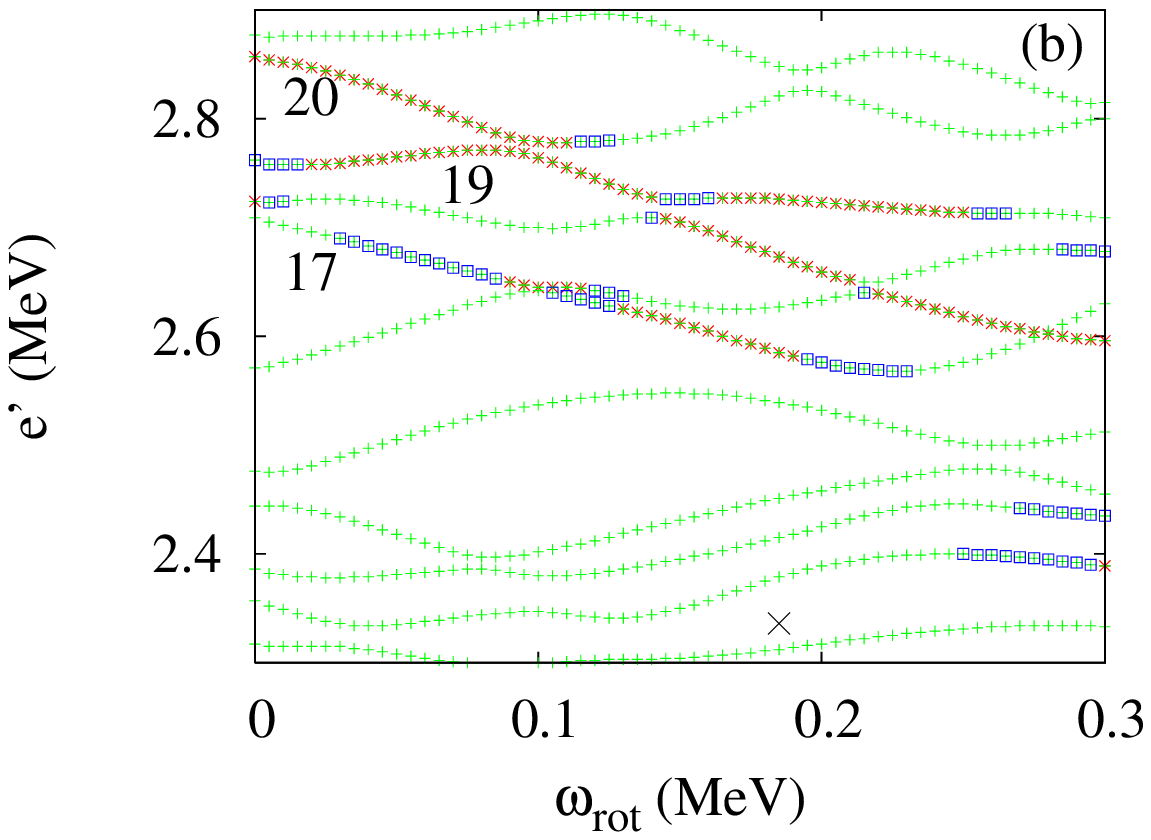}
 \includegraphics[width=6cm]{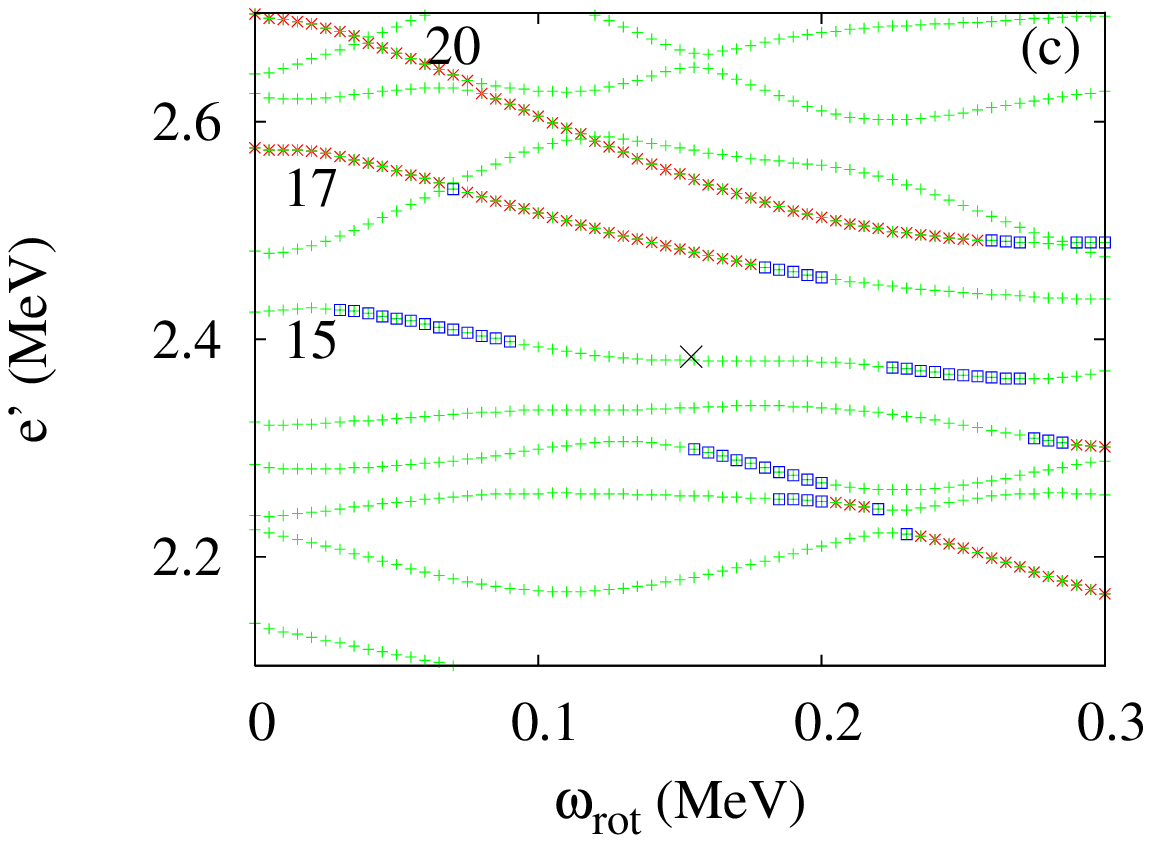}
 \includegraphics[width=6cm]{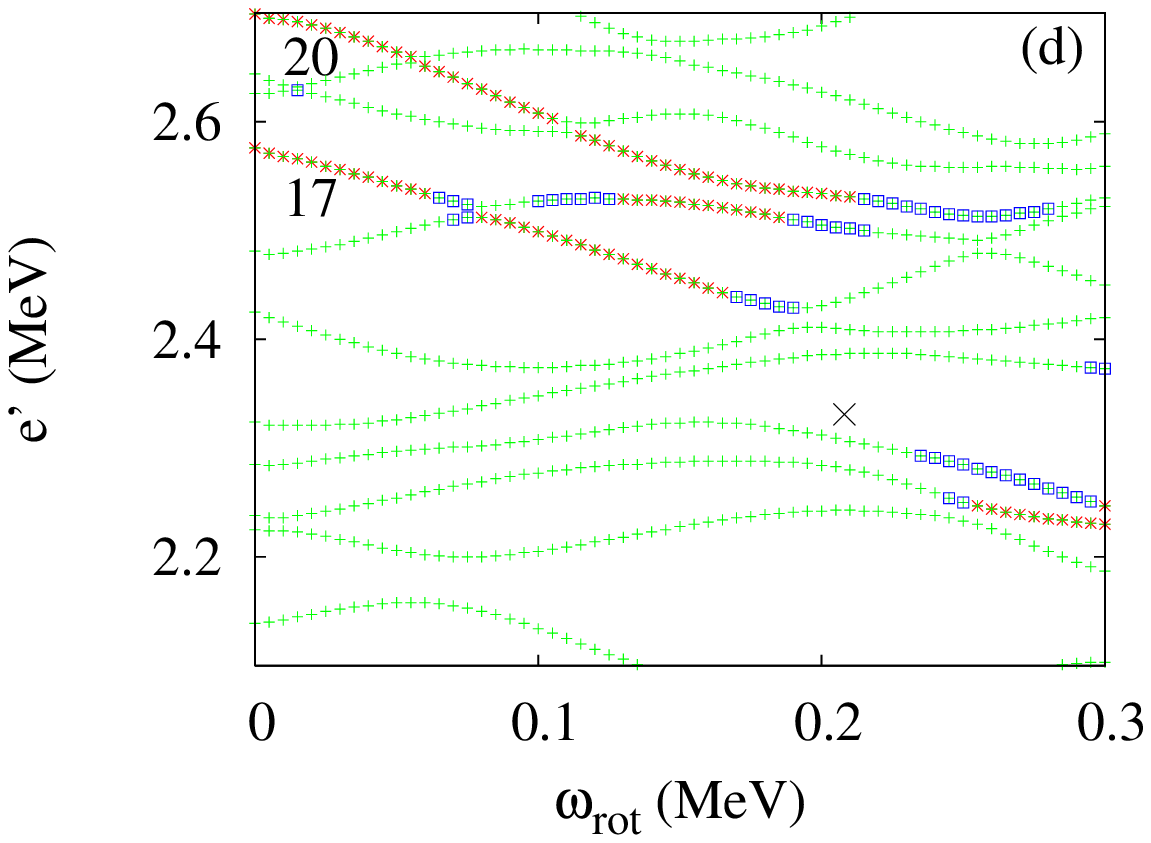}
 \caption{(Color online) Routians of all calculated PVC states in the regions 
of $3\gamma$ states are shown by green $+$s as in Figs.~\ref{fig2} and \ref{fig8}. 
Those with collective fraction more than 50\% and 40 -- 50\% are emphasized 
by red $\times$s and blue $\square$s, respectively. 
(a) $r=-i$ of $^{103}$Nb, (b) $r=+i$ of $^{103}$Nb, (c) $r=-i$ of $^{105}$Nb, 
and (d) $r=+i$ of $^{105}$Nb. The observed transitions in band (4), the $3\gamma$ 
candidate, converted to the rotating frame are also shown by large $\times$s.}
 \label{fig10}
\end{figure}

Calculated eigenstates in the region of $3\gamma$ states are shown in 
Fig.~\ref{fig10}, (a) $^{103}$Nb, favored, (b) unfavored, and (c) $^{105}$Nb, 
favored, (d) unfavored. Contrasting to the previous $1\gamma$ and $2\gamma$ 
cases, collective bands are not always parallel, and collectivity tends to 
move to lower energy states as $\omega_\mathrm{rot}$ increases. 
To show this tendency clearly, states with 40 -- 50\% collective fractions are 
also marked with blue squares in these figures. 

This result can be understood as follows. As discussed already, the 
highest-lying collective states are the most collective and have the 
highest $K$ because of the parallel coupling $\Omega+2n$ for $n\gamma$ states 
at around $\omega_\mathrm{rot}=0$. 
Because of $K\le j_\mathrm{eff}$, where $j_\mathrm{eff}$ is the effective 
single-particle angular momentum of particle-vibration coupled states, 
high-$K$ states have fairly pure high $j_\mathrm{eff}$. 
Consequently they feel strong Coriolis force when rotation sets in. Then 
they start to align their angular momenta to the rotational ($x$) axis with 
reducing $K$, accordingly they reduce their purity and the peak height 
in Figs.~\ref{fig1} and \ref{fig7} gradually. 
This is an aspect of rotational $K$ mixing. Therefore it is 
expected that the collective $3\gamma$ state with $K$ lower than the highest 
value $\Omega+6$ would be observed. Actually, the observed band (4) with $K=9/2$, 
indicated by large crosses in the figures, sits around the location 
determined by connecting the most collective states at $\omega_\mathrm{rot}=0$ 
and 0.3 MeV in the case of $^{105}$Nb. The situation in $^{103}$Nb is to some 
extent similar, 
but (i) the lowest collective state at high $\omega_\mathrm{rot}$ is not 
collective enough and (ii) the calculated states are located higher overall. 
More sophisticated calculation would be desired because these observations 
depend on how band crossings occur and in the present model interband 
interactions occur at the same $\omega_\mathrm{rot}$ rather than the same $I$. 

\subsubsection{Interband $B(E2)$} 

The observed enhanced $B(E2)$s look to be accounted for primarily by vibrational 
collectivity, and the above scenario that the collectivity of the twentieth state 
at $\omega_\mathrm{rot}=0$ is observed with a lower $K$ after band crossing(s) 
can lead to enhanced transitions to the $2\gamma$. Before studying 
$3\gamma\rightarrow2\gamma$ transitions, I check the results of the 
generarized intensity relation on $2\gamma\rightarrow1\gamma$ and 
$1\gamma\rightarrow0\gamma$. 
Some calculated values for $^{103}$Nb that can be compared with the observed 
$B(E2)$ ratios are shown in Table~\ref{table1}. 
Note here that $Q_\mathrm{tr}=0.1944$ eb gives the zero-point amplitude 
$\gamma_0=14^\circ$, defined by $\tan\gamma_0=\frac{\sqrt{2}Q_\mathrm{tr}}{Q_0}$~\cite{BM}, 
by combining with $Q_0=\langle\mathrm{f}|Q_0^{(+)}|\mathrm{f}\rangle=$1.090 eb. 
This is larger than the absolute value of the adopted static triaxial deformation. 

\begin{table}[htbp]
 \caption {Properties of calculated $n\gamma\rightarrow(n-1)\gamma$ transitions with 
$n=$ 1, 2, and 3, designated by the labels of intrinsic states, in $^{103}$Nb. The moment of 
inertia, $\mathcal{J}=32.388$ MeV$^{-1}$, extracted from the energy of the $I=9/2$ 
state in the ground band through $E(I)=\left(I(I+1)-K^2\right)/2\mathcal{J}$, was used.}
 \label{table1}
\begin{center}
\begin{tabular}{rrrrrrrrr} \hline\hline
$r$ & Intr. & $I_\mathrm{i}$ & $K_\mathrm{i}$ & $I_\mathrm{f}$ & $K_\mathrm{f}$ & 
$Q_\mathrm{tr}$ (eb) & $Q_2$ (eb) & $B(E2)$ ($\mathrm{e}^2\mathrm{b}^2$) \\ \hline
$+i$ &  $3\rightarrow1$ & 11/2 &  9/2 &  7/2 &  5/2 & 0.1944 & 0.0044 & 0.02618 \\
$-i$ &  $3\rightarrow1$ & 13/2 &  9/2 &  9/2 &  5/2 & 0.1944 & 0.0043 & 0.01768 \\
$+i$ &  $9\rightarrow3$ & 15/2 & 13/2 & 11/2 &  9/2 & 0.2704 & 0.0067 & 0.06436 \\
$+i$ & $20\rightarrow9$ & 11/2 &  9/2 & 15/2 & 13/2 & 0.3122 & 0.0116 & 0.19168 \\
\hline\hline
\end{tabular}
\end{center}
\end{table}

First, the calculated values in Table~\ref{table1} give 
$\frac{B(E2:15/2^+_{2\gamma}\rightarrow11/2^+_{1\gamma})}{B(E2:11/2^+_{1\gamma}\rightarrow7/2^+_{0\gamma})}=$ 
2.46 and 
$\frac{B(E2:15/2^+_{2\gamma}\rightarrow11/2^+_{1\gamma})}{B(E2:13/2^+_{1\gamma}\rightarrow9/2^+_{0\gamma})}=$ 
3.64, which are very close to the harmonic vibrational values 2.59 and 3.34 
in Ref.~\cite{Nb103}, while the corresponding experimental values are 1.53(16) 
and 3.45(37), respectively. These show that our PVC model describes the 
observed values precisely aside from the fact that the experimental 
$B(E2:11/2^+_{1\gamma}\rightarrow7/2^+_{0\gamma})$ is slightly enhanced. 

Next, the calculated value, 
$\frac{B(E2:11/2^+_{3\gamma}\rightarrow15/2^+_{2\gamma})}{B(E2:11/2^+_{1\gamma}\rightarrow7/2^+_{0\gamma})}=$ 
7.32, is smaller than the corresponding experimental value, 13.5(11), but within a factor of 2. 
The elementary bosonic property, $\left.\hat{a}|n\right\rangle=\sqrt{n}\left.|n-1\right\rangle$, 
is reflected in the intrinsic matrix element, $Q_\mathrm{tr}$, in Table~\ref{table1}, such as 
$\frac{0.3122}{0.1944}=1.61\simeq\sqrt{3}$. This is modified by the angular-momentum 
dependence brought by $Q_2$, and leads to $Q_\mathrm{out}$ in Eq.~(9). Finally $B(E2)$ is 
obtained by multiplying a Clebsch-Gordan coefficient. This means that the calculated 
value contains some of the enhancement from the angular-momentum effect. 

\begin{table}[htbp]
 \caption {The same as Table \ref{table1} but for $^{105}$Nb. The moment of inertia 
$\mathcal{J}=31.831$ MeV$^{-1}$ was used.}
 \label{table2}
\begin{center}
\begin{tabular}{rrrrrrrrr} \hline\hline
$r$ & Intr. & $I_\mathrm{i}$ & $K_\mathrm{i}$ & $I_\mathrm{f}$ & $K_\mathrm{f}$ & 
$Q_\mathrm{tr}$ (eb) & $Q_2$ (eb) & $B(E2)$ ($\mathrm{e}^2\mathrm{b}^2$) \\ \hline
$+i$ &  $3\rightarrow1$ & 11/2 &  9/2 &  7/2 &  5/2 & 0.1928 & 0.0087 & 0.02065 \\
$-i$ &  $3\rightarrow1$ & 13/2 &  9/2 &  9/2 &  5/2 & 0.1928 & 0.0085 & 0.01156 \\
$+i$ &  $8\rightarrow3$ & 15/2 & 13/2 & 11/2 &  9/2 & 0.2674 & 0.0125 & 0.05047 \\
$+i$ & $20\rightarrow8$ & 11/2 &  9/2 & 15/2 & 13/2 & 0.3122 & 0.0185 & 0.22386 \\
$-i$ & $20\rightarrow8$ & 13/2 &  9/2 & 17/2 & 13/2 & 0.3122 & 0.0185 & 0.22501 \\
\hline\hline
\end{tabular}
\end{center}
\end{table}

I proceed to $^{105}$Nb, in which more conspicuous enhancement of $B(E2)$ is observed. 
In this case $\gamma_0$ is $13^\circ$ and again larger than $|\gamma|$. 
The calculated values in Table~\ref{table2} give 
$\frac{B(E2:15/2^+_{2\gamma}\rightarrow11/2^+_{1\gamma})}{B(E2:11/2^+_{1\gamma}\rightarrow7/2^+_{0\gamma})}=$ 
2.44 and 
$\frac{B(E2:15/2^+_{2\gamma}\rightarrow11/2^+_{1\gamma})}{B(E2:13/2^+_{1\gamma}\rightarrow9/2^+_{0\gamma})}=$ 
4.37, while the corresponding experimental values are 2.01(24) and 1.94(25), 
respectively. The calculated values are similar to those of $^{103}$Nb but in the 
experimental ones the magnitude of the denominator is inverted. 

For those from the $3\gamma$ candidates, the calculated values are 
$\frac{B(E2:11/2^+_{3\gamma}\rightarrow15/2^+_{2\gamma})}{B(E2:11/2^+_{1\gamma}\rightarrow7/2^+_{0\gamma})}=$ 
10.8 and 
$\frac{B(E2:13/2^+_{3\gamma}\rightarrow17/2^+_{2\gamma})}{B(E2:13/2^+_{1\gamma}\rightarrow9/2^+_{0\gamma})}=$ 
19.5, while the corresponding experimental values are 27.5(33) and 40.5(55). 
The degree of enhancement increases from $^{103}$Nb both in the calculation and in the data. 
In the calculation, this is caused by a cooperation of increase of the numerator 
and decrease of the denominator. 
By close examination, the difference in $Q_2$, magnified by the angular-momentum factor, 
$I_\mathrm{f}(I_\mathrm{f}+1)-I_\mathrm{i}(I_\mathrm{i}+1)-\Delta K(K_\mathrm{i}+K_\mathrm{f})=6$, 
produces the difference in the $B(E2)$ in the numerator. 
Again the differences in the ratios from the data are within factors of 2 -- 3. 
This indicates that the main mechanism of the enhancement of $B(E2)$ is the vibrational 
collectivity as expected, and the angular-momentum effect is also important for the enhancement. 
Therefore, from the present analyses, it appears promising that the main component of the 
observed band (4) is a three-phonon $\gamma$ vibration although some mixing with 
states that are not included in the present model would be possible. 

\section{Conclusions}

The single- and multi-phonon $\gamma$-vibrational states in $^{103}$Nb 
and $^{105}$Nb have been studied by using the particle-vibration coupling model 
based on the cranking model and the random-phase approximation. The calculations 
have been done in the model space including up to $4\gamma$ basis states. This is an 
extension of the previous calculations including up to $2\gamma$ basis states. 
Analyses of aligned angular momenta and wave functions identified the highest lying among 
the calculated collective states as the most collective 
and of the highest $K=\Omega+2n$ for $n\gamma$ states at $\omega_\mathrm{rot}=0$. 
Then, $1\gamma$ and $2\gamma$ states have been directly identified with the 
observed ones and shown to reproduce the observed spectrum precisely. 
In the case of $3\gamma$ states, the Routhian of the $K=\Omega+6$ state is lowered 
by strong Coriolis force. Because this alignment process reduces $K$, 
the most collective state is expected to be observed as a band with a lower $K$. 
Based on this scenario, the interband $B(E2)$s for $n\gamma\rightarrow(n-1)\gamma$ transitions 
with $n=$ 1, 2, and 3 have been calculated by adopting the method of the 
generalized intensity relation. 
States with $n=1$ and 2 were found to reproduce well the observed ones. The calculated 
$3\gamma\rightarrow2\gamma$ transition rates have accounted for the observed 
enhancement of the transitions from the $3\gamma$ candidates to $2\gamma$ states within 
factors of 2 -- 3, primarily by the vibrational collectivity and secondly by the 
angular-momentum effect. 
These analyses indicate that the main component of the observed band (4) is a 
three-phonon $\gamma$ vibration.

\end{document}